\documentclass[amsmath,amssymb,aps,physrev,reprint,superscriptaddress]{revtex4-2}

\usepackage{amsmath}
\usepackage{amssymb}
\usepackage{mathtools}
\usepackage{graphicx}
\usepackage{dcolumn}
\usepackage{bm}
\usepackage{orcidlink}
\usepackage[capitalize]{cleveref}
\usepackage{rotating}
\usepackage{multirow}
\usepackage{makecell}
\usepackage{enumitem}
\usepackage[FIGTOPCAP]{subfigure}
\usepackage{stackengine}
\usepackage{siunitx}
\usepackage{multirow}
\DeclareMathOperator*{\argmax}{arg\,max}
\DeclareMathOperator*{\argmin}{arg\,min}

\newcommand{\sbullet}[1][.5]{\mathbin{\vcenter{\hbox{\scalebox{#1}{$\bullet$}}}}}

\begin{document}

\preprint{}

\title{Scalability Challenges in Variational Quantum Optimization under Stochastic Noise}


\author{Adelina Bärligea\orcidlink{https://orcid.org/0009-0008-5497-1941}}
\email[]{Contact author: adelina.baerligea@tum.de}
\affiliation{Fraunhofer Institute for Cognitive Systems IKS, Munich, Germany}
\affiliation{Physics Department, Technical University of Munich, Garching, Germany}

\author{Benedikt Poggel\orcidlink{https://orcid.org/0000-0001-9734-3592}}
\affiliation{Fraunhofer Institute for Cognitive Systems IKS, Munich, Germany}

\author{Jeanette Miriam Lorenz\orcidlink{https://orcid.org/0000-0001-6530-1873}}
\affiliation{Fraunhofer Institute for Cognitive Systems IKS, Munich, Germany}
\affiliation{Faculty of Physics, Ludwig-Maximilians-University Munich, Munich, Germany}




\begin{abstract}
With rapid advances in quantum hardware, a central question is whether quantum devices with or without full error correction can outperform classical computers on practically relevant problems. Variational Quantum Algorithms (VQAs) have gained significant attention as promising candidates in this pursuit, particularly for combinatorial optimization problems.
While reports of their challenges and limitations continue to accumulate, many studies remain optimistic based on small-scale, idealized testing setups, leaving doubt about the scalability of VQAs for large-scale problems.
We systematically investigate this scaling behavior by analyzing how classical optimizers minimize variational quantum loss functions for random QUBO instances in the presence of uncertainty, modeled as effective Gaussian noise. 
We find that the critical noise threshold for successful optimization decreases rapidly as system size grows. This decline exceeds what can be explained solely by shrinking loss variance, confirming deeper, fundamental limitations in the loss landscapes of VQAs beyond barren plateaus. Translating these thresholds into required measurement shots reveals that achieving sufficient precision in the evaluated loss values quickly becomes impractical, even for moderately-sized problems.
Our findings demonstrate serious scalability challenges for VQAs in optimization stemming from mere uncertainty, indicating potential barriers to achieving practical quantum advantage with current hybrid approaches.

\end{abstract}


\maketitle
\newpage

\section{\label{sec:introduction}Introduction}

While quantum hardware has advanced rapidly~\cite{Arute.2019,Bruzewicz.2019,Zhong.2020}, scalable and fault-tolerant quantum computing may still be a long way off. John Preskill~\cite{Preskill.2018} coined the term NISQ (Noisy Intermediate-Scale Quantum) to describe a generation of quantum devices, which operate with limited resources and significant noise levels. An open question of the NISQ era is whether these devices can achieve a practically relevant quantum advantage. In this context, variational quantum algorithms (VQAs) have emerged as potentially promising candidates for realizing near-term quantum advantage in a wide range of applications~\cite{McClean.2016,Cerezo.2021,Bharti.2022}. One of the most anticipated use cases, particularly in industry, is combinatorial optimization, where current quantum approaches mainly rely on approximation algorithms and heuristics such as VQAs~\cite{Abbas.2023}.

VQAs are hybrid schemes that leverage quantum computers to evaluate a loss function while relying on classical optimization methods for loss minimization. They employ shallow, parameterized quantum circuits (ansatzes), making them well-suited to the resource constraints of NISQ devices. Yet, a growing body of research highlights significant challenges and limitations of VQAs: One major issue is the barren plateau phenomenon~\cite{McClean.2018,Larocca.2024}, where the loss values and gradients vanish exponentially with system size in most of the proposed quantum ansatzes, making parameter optimization intractable. Conversely, structural properties leading to the provable absence of barren plateaus might enable efficient classical simulations~\cite{Cerezo.2023,Goh.2025}. Additionally, the classical optimization problem within VQAs has been proven to be NP-hard~\cite{Bittel.2021}, with loss landscapes plagued by numerous suboptimal local minima~\cite{Anschuetz.2022}, further complicating the convergence to high-quality solutions. These challenges are amplified by noise, which not only leads to unfavorable resource scaling and unpredictable behavior but may also eliminate any potential quantum advantage altogether~\cite{Leymann.2020,Wang.2021,GonzalezGarcia.2022,Scriva.2024,StilckFranca.2021,Shao.2023,Schuster.2024}. Crucially, these limitations persist regardless of whether the effective noise comes from hardware imperfections or finite sampling errors inherent to quantum measurement. Consequently, despite ongoing efforts and widespread optimism, no practically meaningful quantum advantage has been demonstrated for VQAs on near-term devices to date.

Thus, the critical question remains: can variational quantum optimization become practical in the presence of unavoidable stochastic errors? While some studies directly address this question (see \cref{sec:relatedwork}), most research focuses on developing new heuristic approaches, typically tested on small-scale and idealized setups. Such benchmarks often yield promising insights but rarely capture how well algorithms scale under realistic noise conditions and increasing problem sizes, potentially painting an overly optimistic picture of their viability. More importantly, many studies overlook the role of the classical optimization process -- the central computational bottleneck of VQAs -- which may ultimately determine whether a problem remains solvable as system size grows. Closing this gap is essential for distinguishing genuinely promising advancements from appealing but unproven claims in quantum optimization and related disciplines.

This work investigates the viability of VQAs for solving classical optimization problems in the Quadratic Unconstrained Binary Optimization (QUBO) formulation, with a specific focus on the classical optimization routine as a key driver of performance. We systematically analyze how state-of-the-art classical optimizers handle variational quantum loss functions under increasing levels of Gaussian noise, providing an empirical assessment of practical scalability limits in variational quantum optimization. The primary contributions of this study are:

\begin{enumerate}[leftmargin=*]
    \item \textbf{Performance Evaluation Framework:} We propose a systematic methodology to assess the scaling behavior of hybrid variational quantum optimization methods under stochastic noise, providing a robust foundation for understanding their potential and limitations, even in fault-tolerance.
    \item \textbf{Scaling Analysis of Classical Optimizers:} This study presents, to the best of our knowledge, the first in-depth scaling analysis of classical optimizer performance on noisy quantum loss functions for random QUBO problem instances. By examining two essential factors -- noise level ($\sigma$) and system size ($n$) -- we identify a clear threshold $\sigma^*(n)$, beyond which reliable optimization becomes exponentially challenging. Via extrapolation to regions not (yet) accessible numerically we address the feasibility of these algorithms for larger-scale problems.
    \item \textbf{Insights on Practical Quantum Advantage:} We quantify how uncertainty in the loss values, particularly arising from finite measurement sampling, can significantly impact the practical scalability of VQAs. These results highlight potential barriers to achieving quantum advantage in variational quantum optimization under realistic noise scenarios.
\end{enumerate}

Our empirical findings offer clear benchmarks and valuable insights into the practical limits of VQAs under noise. By fostering a more realistic understanding of the potential of quantum optimization on near-term devices and beyond, these insights help to guide future research towards more promising algorithms and applications in quantum computing.

\section{\label{sec:relatedwork}Related Work}
The trainability of VQAs has been extensively explored in the literature. Several key obstacles impairing successful parameter optimization have been identified, including reachability deficits~\cite{Akshay.2020}, inherent NP-hardness of optimization~\cite{Bittel.2021}, complex loss landscapes with numerous suboptimal local minima~\cite{Anschuetz.2022}, and barren plateaus~\cite{McClean.2018,Larocca.2024}, i.e.~the phenomenon of exponentially vanishing loss values and gradients. Specifically, recent work indicates that all sources of barren plateaus, such as high ansatz expressibility~\cite{McClean.2018}, global observables~\cite{Cerezo.2021b}, or hardware-induced noise~\cite{Wang.2021}, are fundamentally connected to the scaling behavior of the dynamical Lie algebra (DLA) associated with the chosen ansatz~\cite{Cerezo.2021b,Larocca.2022,Fontana.2024,Ragone.2024}. Our study explicitly circumvents these issues by considering only ansatz circuits characterized by only polynomially growing DLA dimensions along with strictly local observables, ensuring that the underlying quantum loss functions are provably barren-plateau-free and remain ``trainable'' in theory. 

A significant body of research has also addressed the detrimental effects of hardware noise on VQA performance. Numerous studies have demonstrated that even minimal noise levels can rapidly propagate through quantum circuits, resulting in output distribution that become effectively simulable classically~\cite{StilckFranca.2021,Bravyi.2021,GonzalezGarcia.2022,Deshpande.2022,DePalma.2023,Shao.2023,Schuster.2024}, thereby negating potential quantum advantage. In contrast, our analysis assumes a fault-tolerant scenario, simulating uncertainty exclusively via Gaussian noise applied as a post-processing step to exact quantum loss function evaluations. This setup effectively models an idealized, fully error-corrected scenario with finite sampling noise as the dominating source of uncertainty. This setting not only circumvents previously identified noise-induced limitations but also aligns more closely with realistic conditions under which quantum advantage for (approximate) optimization can be expected to first emerge.

Given the empirical and experimental nature of our study and its focus on the classical optimization routine, our findings naturally compare to existing numerical assessments of classical optimizer performance within VQAs~\cite{Nakanishi.2019,Sung.2020,Lavrijsen.2020,PellowJarman.2021,Soloviev.2022,Singh.2023,Palackal.2023,BonetMonroig.2023,PellowJarman.2024}. However, as already noted in Ref.~\cite{Barligea.2024}, these studies often evaluate performance under fixed conditions using static setups primarily focusing on comparing optimizer effectiveness. Consequently, they offer limited insights into scalability constraints and the quantitative limits of classical optimizer performance in hybrid quantum algorithms with respect to noise thresholds or system size.

Most closely aligned with our work in terms of scope and metrics is the more recent study by Scriva et.~al.~\cite{Scriva.2024}, which employs similar performance metrics and comes to similar conclusions. However, methodological differences exist: Scriva et.~al.~\cite{Scriva.2024} focus exclusively on shot noise and evaluate iteration complexity relative to fixed success probabilities. Conversely, our approach systematically investigates the effects of varying levels of Gaussian noise, allowing for a broader and more generalizable analysis of the precision requirements for successful optimization. A complementary theoretical investigation into the convergence properties of classical optimizers within a similar context is provided by Kungurtsev et.~al.~\cite{Kungurtsev.2024}.

\section{\label{sec:theory}Theoretical Background}
Optimization has long been regarded as a promising application of quantum computing, particularly for combinatorial optimization problems~\cite{Korte.2018}, where the number of possible solutions grows exponentially with problem size. Many such problems are NP-hard~\cite{Karp.1972,Ausiello.2003}, making them intractable for classical algorithms in the worst case. Quantum algorithms offer several potential approaches to tackling these challenges~\cite{Abbas.2023}, including Grover search~\cite{Grover.1996}, quantum phase estimation~\cite{Nielsen.2016}, adiabatic quantum computing~\cite{Farhi.2000} and heuristic approximation algorithms. Among these, variational quantum algorithms~\cite{Cerezo.2021} have gained particular attention due to their adaptability to near-term quantum devices. 
Although extensively studied~\cite{Bochkarev.2024,Heng.2022,Zahedinejad.2017,Sankar.2024,Dupont.2025}, no practically relevant quantum advantage has been demonstrated in quantum optimization yet. While promising results exist for small-scale or noise-free scenarios, it remains an open question whether these approaches can ultimately outperform classical heuristics, which are already highly optimized for practical applications.

\subsection{\label{ssec:quboproblems}Quadratic Unconstrained Binary Optimization Problems}
A QUBO problem is a widely used mathematical framework for formulating combinatorial optimization problems, defined as
\begin{equation}
    \min_{\mathbf{x}}\ \mathbf{x}^{\top} \mathbf{Q}\: \mathbf{x},
\label{eq:QUBOproblem}
\end{equation}
where $\mathbf{x}\in\{0,1\}^n$ is a vector of binary decision variables, and $\mathbf{Q}\in\mathbb{R}^{n\times n}$ is a real-valued cost matrix that fully characterizes the problem. Due to its versatility, the QUBO formulation can encode a broad range of important optimization problems~\cite{Fu.1986,Kochenberger.2014,Glover.2022,Glover.2022a}. Beyond its flexibility, QUBO is closely linked to statistical physics, specifically to Ising spin glasses~\cite{Barahona.1982}. This connection allows any QUBO problem to be reformulated as an Ising Hamiltonian~\cite{Lucas.2014} (cf.~\cref{appsec:qubo2isingmapping}). This equivalence provides a bridge between classical optimization and quantum physics motivating the application of quantum computing to combinatorial optimization problems. 

The Ising Hamiltonian of an $n$-spin system is given by:
\begin{equation}
    \mathbf{C}=\sum_{i=1}^n \mathcal{C}_{i i} \mathbf{Z}_i+\sum_{i<j} \mathcal{C}_{i j} \mathbf{Z}_i \mathbf{Z}_j,
\label{eq:IsingCostFunction}
\end{equation}
where $\mathbf{Z}_i$ is the Pauli-$\mathbf{Z}$ operator acting on the $i^{\mathrm{th}}$ spin,  with eigenvalues $\pm1$, $\mathcal{C}_{ij}$ (for $i\neq j$) represents the interaction energy between spins $i$ and $j$, and $\mathcal{C}_{ii}$ describes the interaction between each spin with its external magnetic field. Given a computational basis state $|\boldsymbol{q}\rangle = \bigotimes_{i=1}^n|q_i\rangle$ with $q_i\in\{0,1\}$, its total energy ${\langle \mathbf{C}\rangle := \langle\boldsymbol{q}| \mathbf{C}|\boldsymbol{q}\rangle}$ is given by
\begin{equation}
\langle \mathbf{C}\rangle=\sum_{i=1}^n(-1)^{q_i}\: \mathcal{C}_{i i}+\sum_{i<j}(-1)^{q_i+q_j}\: \mathcal{C}_{i j}.
\label{eq:IsingCostExpectation}
\end{equation}
Thus, finding the ground state $|\boldsymbol{q}^*\rangle$ of the system corresponds to minimizing the expectation value of \cref{eq:IsingCostExpectation} with respect to $|\boldsymbol{q}\rangle$, which is known as the Ising problem -- a direct equivalent of the QUBO problem in the optimization context.

\subsection{\label{ssec:vqas}Variational Quantum Optimization}

Until fault-tolerant quantum computing becomes viable, variational quantum algorithms have emerged as a leading heuristic approach for tackling computationally hard problems, such as combinatorial optimization and other complex applications like molecular simulations~\cite{McClean.2016,Cerezo.2021,Bharti.2022}.

VQAs are hybrid quantum-classical algorithms where a quantum computer evaluates the expectation value of a Hermitian cost operator $\mathbf{C}$, for example given by \cref{eq:IsingCostFunction}, with respect to a candidate state
\begin{equation}
|\Psi(\boldsymbol{\theta})\rangle=\mathbf{U}(\boldsymbol{\theta})|\Psi_0\rangle,
\label{eq:candidatestate}
\end{equation}
where $\mathbf{U}(\boldsymbol{\theta})$ is a parameterized quantum circuit (PQC) acting on an easily prepared initial state $|\Psi_0\rangle$, such as the uniform superposition. The classical computer then performs an optimization over the parameters $\boldsymbol{\theta}$ to solve
\begin{equation}
\min_{\boldsymbol{\theta}}\ \mathcal{L}(\boldsymbol{\theta}) \quad\mathrm{with}\quad \mathcal{L}(\boldsymbol{\theta}):=\langle\Psi(\boldsymbol{\theta})|\mathbf{C}| \Psi(\boldsymbol{\theta})\rangle.
\label{eq:parametrizedCostFunction}
\end{equation}
With optimal parameters $\boldsymbol{\theta}^*$, the optimized state $|\Psi(\boldsymbol{\theta^*})\rangle$ approximates the true ground state $|\boldsymbol{q}^*\rangle$. This is a consequence of the variational principle, which ensures that $\langle\psi|\mathbf{C}|\psi\rangle$ is always lower bounded by the ground state energy of the system~\cite{McClean.2016}.

While the classical optimization methods used in VQAs are often well-established, much of the research effort focuses on designing well-behaved quantum ansatzes, which define both the PQC structure and the measurement strategy for estimating the loss function, such as in \cref{eq:IsingCostExpectation}. 
Two of the most widely studied VQAs are the Variational Quantum Eigensolver (VQE)~\cite{Peruzzo.2014}, initially developed for quantum chemistry, and the Quantum Approximate Optimization Algorithm (QAOA)~\cite{Farhi.2014}, a VQE variant explicitly designed for the field of combinatorial optimization. Both algorithms have been extensively explored for a variety of general optimization tasks~\cite{Moll.2018,Nannicini.2019,Tilly.2022,Blekos.2023} since then, making them and their adaptations central candidates in the search for near-term quantum advantage.

\begin{figure*}
    \subfigure{
        \stackon[5pt]{
        \includegraphics[width=0.36\linewidth]{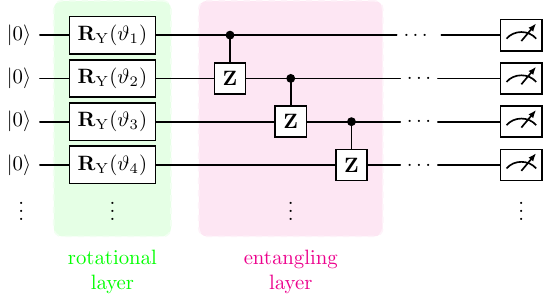}}{
        \textbf{a) Two-Local VQE Ansatz}
        }
    }
    \hfill
    \subfigure{
        \stackon[5pt]{
        \includegraphics[width=0.56\linewidth]{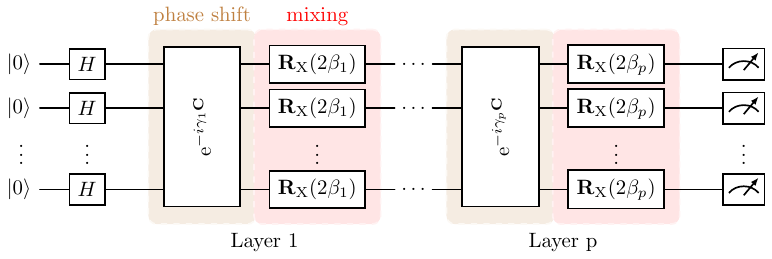}}{
        \textbf{b) Conventional QAOA Ansatz}
        }
    }
    \vspace{-1\baselineskip}
    \caption{\label{fig:algorithms} Representative VQA ansatz circuits. (a) Two-local VQE ansatz. After initializing all qubits to $|0\rangle^{\otimes n}$, alternating layers of singe-quit rotations $\mathbf{R}_{\mathrm{Y}}(\vartheta)$ and linear entangling operations via controlled-$\mathbf{Z}$ gates are applied. A global measurement follows to evaluate the circuit's output. Here, ``two-local'' refers to the ansatz structure, where each entangling gate acts only on pairs of qubits. (b) Conventional QAOA ansatz. After initializing all qubits to the equal superposition state $|+\rangle$ using Hadamard gates, $p$ alternating layers of phase shift and mixing operators are applied as described in \cref{eq:qaoaooperator}, again, followed by a global measurement.}
\end{figure*}

\cref{fig:algorithms} illustrates representative ansatz circuits for both algorithm classes, which are also used in the experiments presented in \cref{ssec:2lossfunctions}. For VQE, a typical example is a hardware-efficient ansatz, consisting of alternating layers of single-qubit rotations and entangling operations, followed by a global measurement. QAOA, in contrast, follows a problem-inspired approach, implementing a trotterized version of adiabatic quantum optimization~\cite{Farhi.2000}. It consists of alternating applications of a shift operator ${\mathbf{U}_P(\gamma)=e^{-i \gamma \mathbf{H}_P}}$ based on the problem Hamiltonian $\mathbf{H}_P$, and a mixing operator ${\mathbf{U}_M(\beta)=e^{-i \beta \mathbf{H}_M}}$, where the default choice is a sum of Pauli-$\mathbf{X}$ matrices on each qubit, i.e., ${\mathbf{H}_M=\sum_{i=1}^n \mathbf{X}_i}$. The resulting QAOA ansatz is composed of $p$ layers, parameterized by $2p$ angles, with the unitary evolution
\begin{equation}
\mathbf{U}(\boldsymbol{\gamma}, \boldsymbol{\beta}) =
\mathbf{U}_M\left(\beta_p\right) \mathbf{U}_P\left(\gamma_p\right) \dots \mathbf{U}_M\left(\beta_1\right) \mathbf{U}_P\left(\gamma_1\right).
\label{eq:qaoaooperator}
\end{equation}
The unitary is applied to an initial state -- typically the uniform superposition -- and the objective is evaluated as in \cref{eq:parametrizedCostFunction}.

A recently proposed alternative to these well-established algorithms is the Block Encoding Quantum Optimizer (BENQO)~\cite{Meli.2023,Bärligea.2024}, which introduces a novel VQA architecture while demonstrating promising results on small-scale problems. Designed explicitly for solving Ising problems, BENQO employs block encoding~\cite{Martyn.2021} -- a technique that embeds the cost operator {$\mathbf{C}\in\mathbb{R}^{2^n\times 2^n}$} \eqref{eq:IsingCostFunction} into a larger unitary matrix at the expense of one additional qubit: 
\begin{equation}
    \mathbf{U}:=\left[\begin{array}{rr}
    \sin (\hat{\mathbf{C}}) & \cos (\hat{\mathbf{C}}) \\
    \cos (\hat{\mathbf{C}}) & -\sin (\hat{\mathbf{C}})
    \end{array}\right] \quad \text{with}\quad \hat{\mathbf{C}}:=\mathbf{C} / K,
\label{eq:BENQOblockencoding}
\end{equation}
where $K$ is a scaling factor that ensures a monotonic relationship between $\langle\mathbf{U}\rangle$ and $\langle\mathbf{C}\rangle$. Then, $\langle\mathbf{U}\rangle$ can be used interchangeably with $\langle\mathbf{C}\rangle$ in the optimization context. Unlike conventional VQAs, which rely on global measurements to estimate expectation values (cf.~\cref{fig:algorithms}), BENQO circumvents this step by using the Hadamard test~\cite{Aharonov.2008}. This method projects the eigenvalues of $\mathbf{C}$ onto an ancillary qubit, allowing for their determination via single-qubit measurements.

\begin{figure}
\includegraphics[scale=0.69]{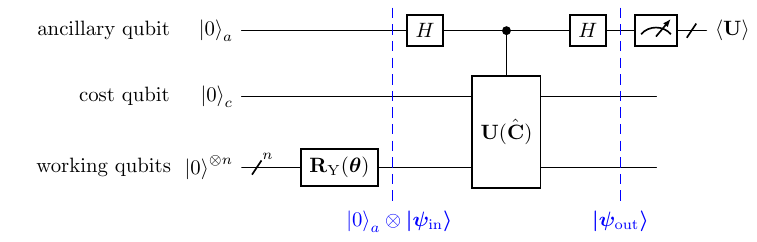}
\caption{\label{fig:benqo} BENQO's ansatz circuit. The quantum system begins in the initial state $|0\rangle_a\otimes|\Psi_{\text{in}}\rangle$ \eqref{eq:benqosinput}, including an ancillary qubit. The Hadamard test then projects the system onto the eigenspace of $\mathbf{U}$ via the projectors $\mathbf{P_{\pm}}$, producing the output state $|\psi_{\text{out}}\rangle$ \eqref{eq:benqosoutput}. Measuring the ancillary qubit yields probabilities $p_0$ and $p_1$, from which the expectation value of $\mathbf{U}$ and ultimately $\mathbf{C}$ can be extracted as in \cref{eq:benqoscost}. This enables direct measurement of the cost function using a single qubit.}
\end{figure}

\cref{fig:benqo} illustrates BENQO's ansatz circuit. The algorithm first applies a rotational layer ${|\mathbf{\psi}(\boldsymbol{\theta})\rangle=\bigotimes_{i=1}^{n}\mathbf{R}_{\mathrm{Y}}(\theta_{i})|0\rangle}$, parameterized by $\boldsymbol{\theta}\in\mathbb{R}^n$, bringing the system into an initial state including an ancillary qubit: 
\begin{equation} 
|0\rangle_a\otimes|\Psi_{\text{in}}\rangle,\quad \mathrm{with}\ |\Psi_{\text{in}}\rangle=|0\rangle_c\otimes|\mathbf{\psi}(\boldsymbol{\theta})\rangle. 
\label{eq:benqosinput}
\end{equation} 
The Hadamard test then projects the system onto the eigenspace of $\mathbf{U}$ using the projectors $\mathbf{P_{\pm}}=\frac{\mathbf{I}_n\pm \mathbf{U}}{2}$. The resulting output state is: 
\begin{equation} {\boldsymbol{|\psi_{\text{out}}\rangle} =|0\rangle_a \otimes\left(\mathbf{P_{+}}|\Psi_{\text{in}}\rangle\right) +|1\rangle_a \otimes\left(\mathbf{P_{-}}|\Psi_{\text{in}}\rangle\right)}. 
\label{eq:benqosoutput}
\end{equation}
Measuring the ancillary qubit allows BENQO to extract the expectation value of $\mathbf{U}$ via the probabilities: \begin{equation} p_0 = \lVert \mathbf{P_{+}}|\psi_{\text{in}}\rangle \rVert^2, \quad p_1 = \lVert \mathbf{P_{-}}|\psi_{\text{in}}\rangle \rVert^2. \end{equation} From this, the cost function is obtained as ${\langle\mathbf{U}\rangle\equiv p_0 - p_1}$, leading to 
\begin{equation} 
\langle\mathbf{C}\rangle=K\arcsin(p_0-p_1).
\label{eq:benqoscost}
\end{equation}
This unique measurement strategy distinguishes BENQO from conventional VQAs, providing an alternative route for quantum optimization. To showcase that the benchmarking and scaling analysis is broadly applicable beyond standard VQA methods, BENQO is included in the comparative loss function studies in \cref{ssec:2lossfunctions}.

\section{\label{sec:methods}Experimental Setup}

\begin{figure*}
\includegraphics[width=0.9\linewidth]{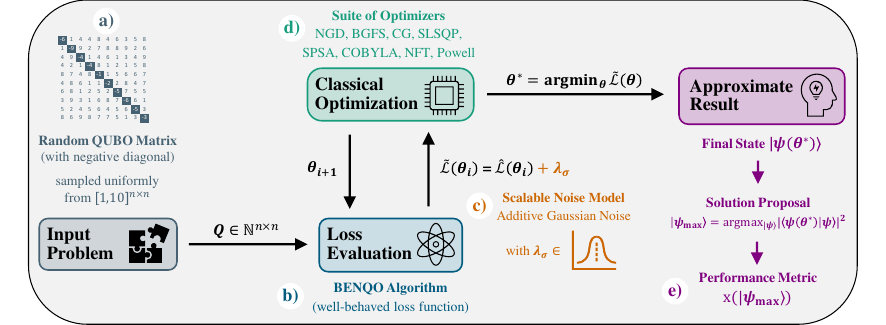}
\caption{\label{fig:methodology} Schematic overview of the experimental setup. (a) A random QUBO problem is sampled as input to the VQA pipeline (\cref{ssec:1probleminstance}). (b) The BENQO algorithm is simulated to obtain exact loss values for the given problem, chosen for its particularly well-behaved loss landscape (\cref{ssec:2lossfunctions}). (c) The exact loss values are perturbed by additive Gaussian noise before being passed to a classical optimizer (\cref{ssec:3noisemodel}). (d) A suite of well-established classical optimization methods (\cref{ssec:4optimizers}) attempts to find the optimal parameters minimizing the noisy loss function. (e) The results are then evaluated based on a predefined success criterion (\cref{ssec:5performancemetrics}).}
\end{figure*}

To ensure a generalizable scaling analysis, we account for the fact that the three main components of VQAs -- problem encoding, the quantum subroutine (for loss evaluation), and the classical subroutine (for optimization) -- each contribute differently to overall performance. To isolate the impact of classical optimization, the central component of VQAs, we fix the other two components under representative conditions, ensuring that our analysis directly reflects the capabilities of parameter optimizers rather than external constraints. 
\cref{fig:methodology} provides a schematic overview of the experimental setup:

\begin{enumerate}[leftmargin=*]
    \item \textbf{Problem Encoding:} Random QUBO instances are used with the additional stipulation that the diagonal of the QUBO matrices contains only negative values. QUBO can be regarded as the standard formulation for combinatorial quantum optimization in the VQA context and provides a realistic benchmark (cf.~\cref{ssec:1probleminstance}).
    \item \textbf{Quantum Subroutine:} The loss function is derived from BENQO, as it exhibits a particularly well-behaved loss landscape compared to other established methods (cf.~\cref{ssec:2lossfunctions}).
    \item \textbf{Noise Modelling:} Instead of relying on hardware-specific noise models, we introduce effective Gaussian noise directly to the loss function, ensuring a highly generalizable and systematic approach to model stochastic errors (cf.~\cref{ssec:3noisemodel}) regardless of their origin.
    \item \textbf{Classical Subroutine:} To eliminate biases arising from specific optimizer choices, we include a full suite of state-of-the-art methods in our experiments (cf.~\cref{ssec:4optimizers}).
    \item \textbf{Statistical Evaluation:} By assigning a success index to each optimization run and repeating experiments $100$ times, we derive statistical measures of problem solvability, enabling a systematic scaling analysis (cf.~\cref{ssec:5performancemetrics}).
\end{enumerate}
This structured approach ensures that our findings are not dependent on one specific problem, optimizer, or noise type, allowing us to draw broader conclusions about the scalability of variational quantum optimization. The following sections provide the technical details behind each of these considerations.

\subsection{\label{ssec:1probleminstance}Random Problem Instances}

Rather than focusing on a particular graph problem and mapping it to its Ising equivalent~\cite{Lucas.2014}, this study employs randomly generated problem instances for the experiments. Each instance is uniquely seeded, with all entries of the $n{\times}n$ QUBO matrix $\mathbf{Q}$ sampled from a discrete uniform distribution over the interval $[1,10]$:
\begin{equation}
Q_{i j} \sim 
\begin{cases} 
+\:\mathcal{U}(1,10) & \text{if } i \neq j \\
-\:\mathcal{U}(1,10) & \text{if } i = j
\end{cases},\quad \forall i,j\in\{1,...,n\}.
\label{eq:randomqubomatrix}
\end{equation}
Since many practically relevant problems involve constraints that translate into quadratic penalty terms~\cite{Glover.2022}, the resulting QUBO matrices often feature negative diagonal elements (cf.~\cref{appsec:tspconstraintsexample}). To better reflect this structure, we also impose negative terms in $\mathbf{Q}$ \eqref{eq:randomqubomatrix}. Setting the sampling range to start at $1$ instead of $0$ further ensures that the matrices represent fully connected graph problems, including self-connections. Lastly, the exact magnitude or precision of the weights is not critical, as the resulting loss values are normalized across all experiments.

\subsection{\label{ssec:2lossfunctions}Well-Behaved Quantum Loss Functions}

Trainability in VQA architectures is generally hindered by two key challenges: the barren plateau (BP) phenomenon, characterized by an exponential concentration of the loss function and its gradients~\cite{Larocca.2024}, and other unfavorable loss landscape structures which complicate optimization~\cite{Bittel.2021,Anschuetz.2022}. While empirical studies on the latter are computationally demanding due to the high dimensionality of the parameter space~\cite{Rajakumar.2024}, BPs can be detected more efficiently by analyzing the sample variance of the quantum loss function~\cite{McClean.2018,Larocca.2022}.

Consider the exact loss function $\mathcal{L}(\boldsymbol{\theta})$, parameterized by $\boldsymbol{\theta}\in\mathbb{R}^n$. To ensure consistent quantitative comparisons of the loss variance across different system sizes, this function must be normalized, as suggested in~\cite{Larocca.2022}:
\begin{equation}
    \hat{\mathcal{L}}(\boldsymbol{\theta})=\frac{\mathcal{L}(\boldsymbol{\theta})}{\lvert\mathcal{L}(\boldsymbol{\theta})\rvert_{\max}}\quad \mathrm{with} \quad \lvert\mathcal{L}(\boldsymbol{\theta})\rvert_{\max} = \max_{\boldsymbol{\theta}\in\mathbb{R}^n}\ \lvert\mathcal{L}(\boldsymbol{\theta})\rvert.
\label{eq:normalizedlossfunction}
\end{equation}
This normalization guarantees that the loss landscape is bounded within $[-1,1]$, independent of the problem size $n$. The sample variance can thus be calculated as
\begin{equation}
    Var[\{\mathcal{\hat{L}}_1,...\mathcal{\hat{L}}_N\}]=\frac{1}{N-1}\sum_{i=1}^{N}(\mathcal{\hat{L}}_i-\overline{\mathcal{\hat{L}}})^2,
\label{eq:samplevariance}
\end{equation}
using a set of normalized values $\{\mathcal{\hat{L}}_i{=}\mathcal{\hat{L}}(\boldsymbol{\theta}_i)\}_{i=1,...,N}$ sampled uniformly across the parameter space. 

\begin{figure}
\includegraphics[scale=0.69]{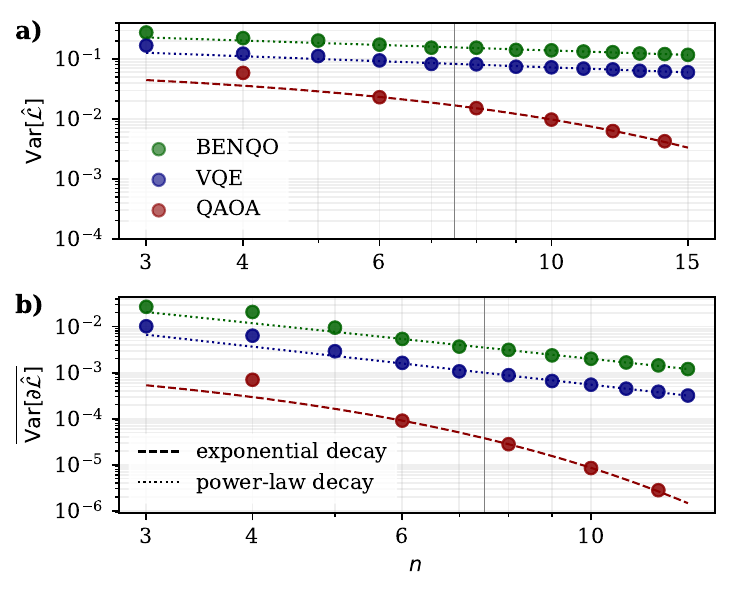}
\caption{\label{fig:varianceanalysis} Variance Analysis of BENQO, VQE and QAOA. (a) Sample variance of the normalized loss values $\mathcal{\hat{L}}(\boldsymbol{\theta})$ computed across $10000$ parameter sets $\boldsymbol{\theta}$, uniformly sampled within $[-2\pi,2\pi]^n$, for increasing system sizes $n$. (b) Mean sample variance of the $n$ partial derivatives $\partial_{\boldsymbol{\theta}}\mathcal{\hat{L}}(\boldsymbol{\theta})$, calculated as \cref{eq:parametershiftrule}, and evaluated at the same $10000$ parameter sets. Both plots include fitted theoretical curves representing exponential and power-law decay, with fitting performed only for $n\geq8$ (indicated by the vertical line) to reduce the influence of small-scale effects.}
\end{figure}

In \cref{fig:varianceanalysis}, we illustrate the scaling of variances across three different VQA ansatzes: a conventional QAOA implementation from \textsc{Qiskit}, a hardware-efficient two-local VQE ansatz, consisting of one rotational layer and one linear entangling layer, and the BENQO algorithm (cf.~\cref{ssec:vqas}). Each ansatz is parameterized such that the number of parameters matches the problem size $n$ \footnote{This is a conservative assumption on parameter scaling, as larger problem instances typically require deeper circuits with more parameters, leading to higher-dimensional loss landscapes and potentially better optimization conditions~\cite{Larocca.2023}. Fixing this relationship therefore ensures a fair scalability comparison without biases from varying parameter growth.}, ensuring a fair and unbiased comparison. Specifically, for the QAOA ansatz this amounts to $p=n/2$ layers, and therefore requires $n$ to be even. 

Based on a characterization of their dynamical Lie algebra (DLA) structures, we know neither the Two Local ansatz used here nor the problem-dependent BENQO ansatz exhibit exponentially scaling DLA dimensions~\cite{Wiersema.2024,Kökcü.2024}. Conversely, QAOA when applied to the Ising cost function defined in \cref{eq:IsingCostFunction} is known to have an exponentially scaling DLA dimension and therefore exhibits a barren plateau~\cite{Larocca.2022,Wiersema.2024}. While various mitigation strategies, such as smart parameter initialization methods (discussed in detail in \cref{ssec:ansatzchoiceinitializations}), can substantially improve the trainability of such ansatzes, this study explicitly concentrates on ansatzes that inherently avoid barren plateaus. Among the two remaining ones tested, BENQO consistently maintains the highest absolute variance in its loss values with increasing system size, suggesting the best parameter sensitivity. Therefore, BENQO is selected for the subsequent experiments as a representative scenario for a favorable case for classical optimization.

\subsection{\label{ssec:3noisemodel}Scalable Gaussian Noise Model}

In our experiments, effective noise is introduced as a post-processing step before the evaluated loss values are passed to the classical optimizers. This ensures that the original, simulated quantum output remains unchanged (cf.~\cref{fig:methodology}).
Since parameter optimizers treat the quantum expectation value as black-box function, they respond only to the total error magnitude rather than its specific source, it is reasonable to approximate the cumulative effect of various noise sources using additive Gaussian noise, as motivated in \cref{appsec:FSerrordistribution} (cf.~\cite{Barligea.2024}). The noise model is applied as follows:
\begin{equation}
    \tilde{\mathcal{L}}_{\sigma}(\boldsymbol{\theta})=\hat{\mathcal{L}}(\boldsymbol{\theta})+\lambda_{\sigma}\quad\mathrm{with}\quad \lambda_{\sigma}\sim\mathcal{N}(0,\sigma).
\label{eq:additivegaussiannoise}
\end{equation}
Here, $\hat{\mathcal{L}}(\boldsymbol{\theta})$ is the normalized exact loss value \eqref{eq:normalizedlossfunction}, and $\lambda_{\sigma}$ is a random variable drawn from a normal distribution with mean $0$ and standard deviation $\sigma$. 

Although this method does not explicitly capture hardware-level error propagation, as explored in other studies~\cite{Sung.2020,Lavrijsen.2020,GonzalezGarcia.2022}, it may offer several advantages: First, it is highly scalable, allowing systematic evaluations across different problem sizes without the computational overhead of full quantum noise simulations. Second, it provides a generalizable and broadly applicable framework by collapsing all potential noise sources into a single tunable noise parameter $\sigma$, independent of the chosen quantum ansatz. As a result, other studies can map their findings onto a comparable $\sigma$ value by computing the standard deviation of their total error, enabling direct comparison regardless of the specific hardware or noise model used.

\subsection{\label{ssec:4optimizers}State-of-the-Art Optimizers}

In VQAs, the principal optimization routine is handled by classical optimizers, which can be broadly classified into gradient-based and gradient-free methods. Gradient-based approaches are often preferred for their quick convergence in practice, but calculating derivatives can be computationally expensive or even infeasible in some cases. Consequently, gradient-free methods are widely used when gradients are unreliable, impractical, or costly to compute. 

This study includes both categories of optimizers, ensuring a diverse representation of optimization strategies. A selection of well-established methods from \textsc{Qiskit}~\cite{Qiskit} and \textsc{SciPy}~\cite{Scipy1.0} as chosen (cf.~\cref{tab:testedoptimizers}), along with a normalized gradient descent method (NGD)~\cite{Murray.2019,Suzuki.2021}, proposed as such by Kuete Meli et al.~\cite{Meli.2023}, which has demonstrated promising performance in prior studies~\cite{Suzuki.2021,Meli.2023,Bärligea.2024}. For the Qiskit optimizers, one should refer to the documentation and references listed in \cref{tab:testedoptimizers}.

\begin{table*}
    \begin{tabular}{c|l|l|l|l}
        & \textbf{Name}&\textbf{Description}& \textbf{Reference} &\textbf{in} \textsc{Qiskit} \\ \hline
        \multirow{4}{*}{\begin{turn}{90}\makecell{gradient-\\based}\end{turn}} & NGD & normalized gradient descent method & Kuete Meli et al.~\cite{Meli.2023} &\textit{custom} \\
        & BFGS & Broyden-Fletcher-Goldfarb-Shannon algorithm & Byrd et al.~\cite{Byrd.1995}& \texttt{L\_BFGS\_B}\\
        & CG & conjugate gradient method & Fletcher~\cite{Fletcher.1964} & \texttt{CG} \\
        & SLSQP & sequential least squares programming method & Kraft~\cite{Kraft.1988} & \texttt{SLSQP} \\ \hline
        \multirow{4}{*}{\begin{turn}{90}\makecell{gradient-\\free}\end{turn}} & SPSA & simultaneous perturbation stochastic approximation & Spall~\cite{Spall.1992,spall.1998} & \texttt{SPSA} \\
        & COBYLA & constrained optimization by linear approximation & Powell~\cite{Powell.1998,Powell.1994} & \texttt{COBYLA} \\
        & NFT & Nakanishi-Fuji-Todo algorithm  & Nakanishi et al.~\cite{Nakanishi.2019} & \texttt{NFT}\\
        & Powell & conjugate direction method & Powell~\cite{Powell.1964} &\texttt{POWELL}
    \end{tabular}
    \caption{\label{tab:testedoptimizers}Overview of tested optimizers. A representative set of both gradient-based and gradient-free parameter optimizers was selected based on a pre-screening process to ensure the inclusion of methods that demonstrated promising performance and efficiency. Their respective class names in \textsc{Qiskit} are shown in the right column.}
\end{table*}

NGD uses a normalized gradient as the search direction and an exponentially decaying step size for its update rule, ensuring stable convergence:
\begin{equation}
    \boldsymbol{\theta}_{k+1}\leftarrow\boldsymbol{\theta}_k- \sqrt{\frac{\pi}{2}\:n}\ \exp \left(-\frac{4 k^2}{k_{\max }^2}\right)\ \frac{\nabla_{\boldsymbol{\theta}}\mathcal{L}(\boldsymbol{\theta}_k)}{\lVert \nabla_{\boldsymbol{\theta}}\mathcal{L}(\boldsymbol{\theta}_k) \rVert_2}.
\end{equation}
Here, the only hyperparameter is the total number of optimization steps, $k_{\max}$, set to $k_{\max}=20$, as recommended in previous studies~\cite{Meli.2023}.

Note that for optimizers requiring exact gradients ${\nabla_{\boldsymbol{\theta}} \mathcal{L}(\boldsymbol{\theta})=\left(\partial_{\theta_1} \mathcal{L}(\boldsymbol{\theta}), \ldots, \partial_{\theta_n} \mathcal{L}(\boldsymbol{\theta})\right)^{\top}}$, the parameter-shift rule~\cite{Mitarai.2018,Schuld.2019} was employed:
\begin{equation}
\frac{\partial}{\partial \theta_i} \mathcal{L}(\boldsymbol{\theta})= \frac{\mathcal{L}(\boldsymbol{\theta}_{i,+})-\mathcal{L}(\boldsymbol{\theta}_{i,-})}{2},
\label{eq:parametershiftrule}
\end{equation}
where $\boldsymbol{\theta}_{i,\pm}:= (\dots, \theta_i\pm\frac{\pi}{2},\dots)$. This exact calculation method requires $2n$ function evaluations per gradient computation, and a total of $(2n+n)k_{\max}$ evaluations for the full optimization.

To ensure a fair comparison among all optimizers, we fix their available computational resources, using default settings from \textsc{Qiskit}. The initial parameters for each optimization run are sampled from a multivariate standard normal distribution centered at $\mathbf{0}\in\mathbb{R}^n$, a strategy shown to be effective in previous research~\cite{Zhang.2022b,Meli.2023,Bärligea.2024}. Each experiment consists of a single optimization run with a fixed optimizer (from \cref{tab:testedoptimizers}), a specific noise level $\sigma$, and a random QUBO problem of size $n$. To account for statistical variation, we repeat each experiment $N=100$ times with different problem instances.

\subsection{\label{ssec:5performancemetrics}Measures of Solvability}
The solution of a variational optimization task is defined as the $n$-dimensional parameter vector 
\begin{equation}
    \boldsymbol{\theta}^*=\argmin_{\boldsymbol{\theta}}\mathcal{L}(\boldsymbol{\theta}),
\label{eq:solutionparameters}
\end{equation}
which (approximately) minimizes the quantum loss function $\mathcal{L}(\boldsymbol{\theta})$. A widely used metric to evaluate the quality of this solution is the approximation ratio (AR),
\begin{equation}
    Q_{\mathrm{AR}}(|\boldsymbol{\psi}^*\rangle) = \frac{\langle\boldsymbol{\psi}^*|\mathbf{C}|\boldsymbol{\psi}^*\rangle-C_{\text{max}}}{C_{\text{min}}-C_{\text{max}}},\quad  Q_{\mathrm{AR}}\in\ [0,1],
\label{eq:approximationratio}
\end{equation}
which quantifies how close the solution state ${|\boldsymbol{\psi}^*\rangle = |\boldsymbol{\psi}(\boldsymbol{\theta}^*)\rangle}$ is to the optimal state $|\boldsymbol{q}^*\rangle$, based on the energy of the underlying quantum system. Here, $\mathbf{C}$ is the Hamiltonian cost operator \eqref{eq:IsingCostFunction}, $C_{\text{min}}=\langle \boldsymbol{q}^*|\mathbf{C}|\boldsymbol{q}^*\rangle$ represents the minimum energy, and $C_{\text{max}}$ is the maximum, according to \cref{eq:IsingCostExpectation}:
\begin{equation}
    C_{\text{max}}= \sum_{i=1}^n\left|\mathcal{C}_{i i}\right|+\sum_{ i<j }\left|\mathcal{C}_{i j}\right|.
\end{equation}

For classical optimization problems, the optimal state $|q^*\rangle$ corresponds to a computational basis state and can be expressed as a binary string. Thus, instead of directly evaluating  $|\boldsymbol{\psi}^*\rangle$ using the AR, the most probable basis state of the quantum output,
\begin{equation}
    |\psi_{\text{max}}\rangle = \argmax_{\psi_i} |\langle\boldsymbol{\psi}^*|\psi_i\rangle|^2,
\label{eq:mostprobablebasisstate}
\end{equation} 
should be considered the solution candidate of the VQA. This choice is motivated by the practical perspective that real-world users aim to directly sample a high-quality solution, rather than undertake the impractical effort of searching for one within a delocalized quantum output distribution. Note that in a real experiment, $|\psi_{\text{max}}\rangle$ would be chosen based on the basis state with the highest measured shot count. However, since we are working with simulated results, we can use the exact definition.

To provide a more representative performance indicator, we use the approximation index $x_t$~\cite{Meli.2023,Bärligea.2024}, which assesses whether the loss value of $|\psi_{\text{max}}\rangle$ is within a specified threshold $t \in [0,1]$ of the global minimum:
\begin{equation}
    x_t(|\psi_{\text{max}}\rangle) = 
    \begin{cases}
    1, & \text{if } Q_{\mathrm{AR}}(|\psi_{\text{max}}\rangle) \geq t \\
    0, & \text{otherwise}.
\end{cases}
\label{eq:approximationindex}
\end{equation}
For example, ${x_1 = 1}$, if and only if the most probable basis state $|\psi_{\text{max}}\rangle$ is the exact optimal solution $|q^*\rangle$. 

Using this index, the probability of successfully finding an optimal (or near-optimal) solution -- referred to as the ``solvability'' -- is estimated as
\begin{equation}
\hat{p}_t = \frac{1}{N} \sum_{i=0}^N x_t(|\psi_{\text{max}}\rangle_i),\quad \hat{p}_t \in\ [0,1],
\label{eq:optimalsolutionprobability}
\end{equation}
where $N$ is the total number of optimization runs. This probability reflects the fraction of runs among all experiments that yielded a solution within the specified threshold $t$. In the results presented in \cref{sec:results}, we denote the probability of finding the exact optimal solutions (${t=1}$) as $\hat{p}_{\mathrm{opt}}$, while near-optimal solutions within a $5\%$ and $1\%$ margin of the optimal are monitored as $\hat{p}_{95\%}$ and $\hat{p}_{99\%}$, respectively. 

The standard error of the sample success probability is given by
\begin{equation}
    \sigma( \hat{p}_t )= \sqrt{\frac{\hat{p}_t(1-\hat{p}_t)}{N}},
\label{eq:Deltaoptimalsolutionprobability}
\end{equation}
which follows from the normal approximation of a binomial distribution and is valid for sufficiently large $N$.

\section{\label{sec:results}Results of Scaling Analysis}
\begin{figure*}
\includegraphics[width=1\linewidth]{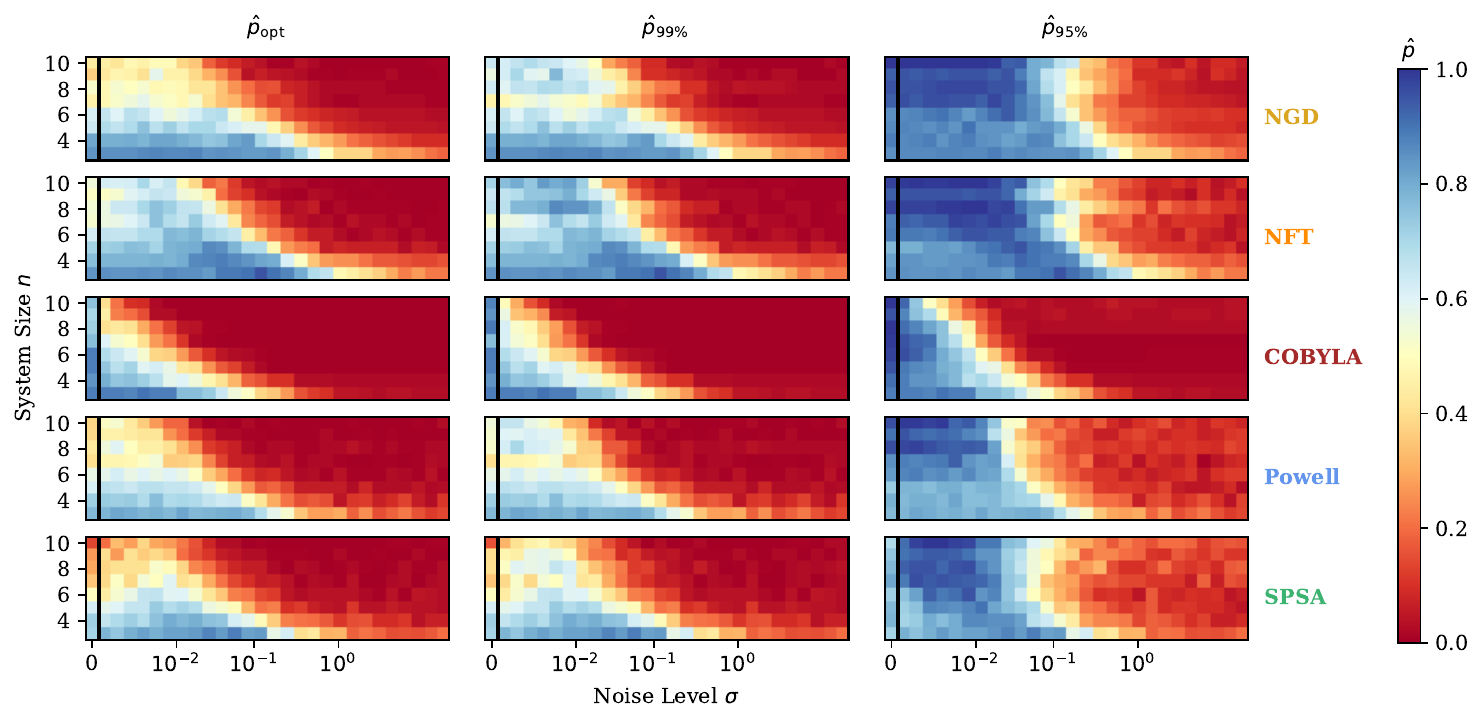}
\caption{\label{fig:gridplot} Success probability across noise levels and system sizes. Contour plots show the proportions $\hat{p}_{\mathrm{opt}}$, $\hat{p}_{99\%}$, and $\hat{p}_{95\%}$ for varying noise levels $\sigma$ (x-axis, logarithmic scale) and system sizes $n$ (y-axis) of the tested optimizers. The leftmost column in each grid represents the no-noise results, which are displayed outside the logarithmic scale for completeness.}
\end{figure*}

The experiments described above are conducted for system sizes ${n\in[3,10]}$ and Gaussian noise levels ${\sigma\in(10^{-3},10^{1})}$. \cref{fig:gridplot} presents the combined results as two-dimensional grids, showing the measured success probabilities $\hat{p}_{\mathrm{opt}}$, $\hat{p}_{99\%}$, and $\hat{p}_{95\%}$ as contours. Note that the optimizers BFGS, CG, and SLSQP are excluded due to their poor performance across all tested noise settings, and their results are therefore omitted from further analysis.

The visualizations in \cref{fig:gridplot} reveal both optimizer-specific differences and common trends across all methods. In particular, a well-defined transition emerges between regions where optimization is successful and those where the success probability drops rapidly to near zero. For exact optimality ($\hat{p}_{\mathrm{opt}}$), this separation appears as a line, while for $\hat{p}_{95\%}$, the boundary takes on a different shape. The following sections provide a detailed quantitative analysis of these observations. 

\subsection{\label{ssec:scalingnoise}Solvability under Additive Gaussian Noise}

\begin{figure}
\includegraphics[scale=0.69]{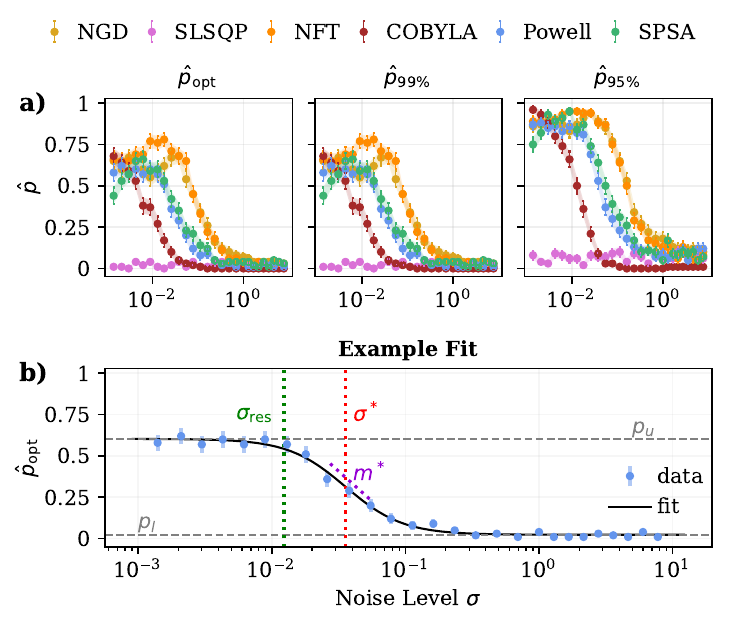}
\caption{\label{fig:scalingnoiselevel} (a) Scaling of success probabilities with increasing noise level. The proportions of optimal ($\hat{p}_{\mathrm{opt}}$) and near-optimal ($\hat{p}_{99\%}$, $\hat{p}_{95\%}$) solutions for problems of size ${n=60}$ are plotted against the noise levels $\sigma$ (logarithmic scale) for a suite of classical optimizers. (b) An exemplary $\tanh$-fit of $\hat{p}_{\mathrm{opt}}$ for the Powell optimizer is shown, with the parameters of noise resilience, $\sigma^*$ \eqref{eq:sigmastar}, $m^*$ \eqref{eq:mstar}, and $\sigma_{\mathrm{res}}$ \eqref{eq:sigmares}, indicated.}
\end{figure}

To analyze the impact of increasing noise levels on solvability, the first scaling analysis is conducted at fixed system size. \cref{fig:scalingnoiselevel} shows the experimental data for an exemplary system size of $n=6$. On a logarithmic scale in the noise level $\sigma$, the curves resemble sigmoidal functions with horizontal asymptotes for $\log(\sigma)\rightarrow\pm\infty$.
As a typical sigmoidal function, the hyperbolic tangent defined as $\tanh(x)=\frac{e^x-e^{-x}}{e^x+e^{-x}}$ is used to fit the measured success probability curves. For easier interpretation, the following form is used:
\begin{equation}
p_{\mathrm{fit}}(\sigma) = \frac{p_u-p_l}{2}\: \tanh\left(-b\log(\sigma)+c\right)+\frac{p_u+p_l}{2}.
\label{eq:tanhfitfunction}
\end{equation}
Here, $p_u$ and $p_l$ represent the upper and lower asymptotes of performance, respectively, and are defined as:
\begin{equation}
    \lim_{\sigma\rightarrow+\infty}\ p_{\mathrm{fit}}(\sigma)=p_l, \qquad
    \lim_{\sigma\rightarrow0}\ p_{\mathrm{fit}}(\sigma)=p_u.
\end{equation}
The parameters $b$ and $c$ quantify the gradient and location of the performance decrease on the logarithmic scale, and offer insights into the noise resilience of the tested solver.
It is noteworthy, however, that the curves of the NFT and SPSA optimizer deviate significantly from a purely sigmoidal shape. Instead, both optimizers exhibit a clear peak in performance at a non-zero noise level before their success probabilities decline. This suggests that a small level of noise can enhance their ability to find optimal solutions by helping to escape saddle points and local minima, a phenomenon also reported in the literature~\cite{Branke.2003,Stich.2021,Liu.2023,Kaminishi.2024}.

To characterize the scaling behavior of the solvability measure $\hat{p}$, several metrics were derived to describe the noise resilience of classical optimizers based on the fit parameters: 
\begin{enumerate}[leftmargin=*]
    \item The \textit{point of steepest descent}, $\sigma^*$, indicates the location of the performance drop and computes to
    \begin{equation}
    \sigma^*=\exp{\left(\frac{c}{b}\right)}.
    \label{eq:sigmastar}
    \end{equation}
    
    \item The \textit{slope at steepest descent}, $m^*$, quantifies the absolute performance decrease at $\sigma=\sigma^*$ and corresponds to the derivative of $p_{\mathrm{fit}}(\sigma)$ at that point:
    \begin{equation}
    m^*=\frac{b\:(p_l-p_u)}{2}\exp\left(-\frac{c}{b}\right).
    \label{eq:mstar}
    \end{equation}
    
    \item The \textit{point of resilience}, $\sigma_{\mathrm{res}}$, serves as another indicator of noise tolerance, defined as the noise level at which the optimizer's performance drops to $90\%$ of its no noise level $p_u$:
    \begin{equation}
    \sigma_{\mathrm{res}}=p_{\mathrm{fit}}^{-1}(0.9\:p_u).
    \label{eq:sigmares}
    \end{equation}
    Here, the $90\%$ threshold is an arbitrary choice made for consistency across experiments.
\end{enumerate}
These metrics offer valuable insights into the onset and rate of performance loss experienced by classical optimizers as noise levels increase. Their behavior across different problem sizes is systematically analyzed in the following section.

\subsection{\label{ssec:scalingsystemsize}Solvability Depending on System Size}

\begin{figure*}
\includegraphics[width=1\linewidth]{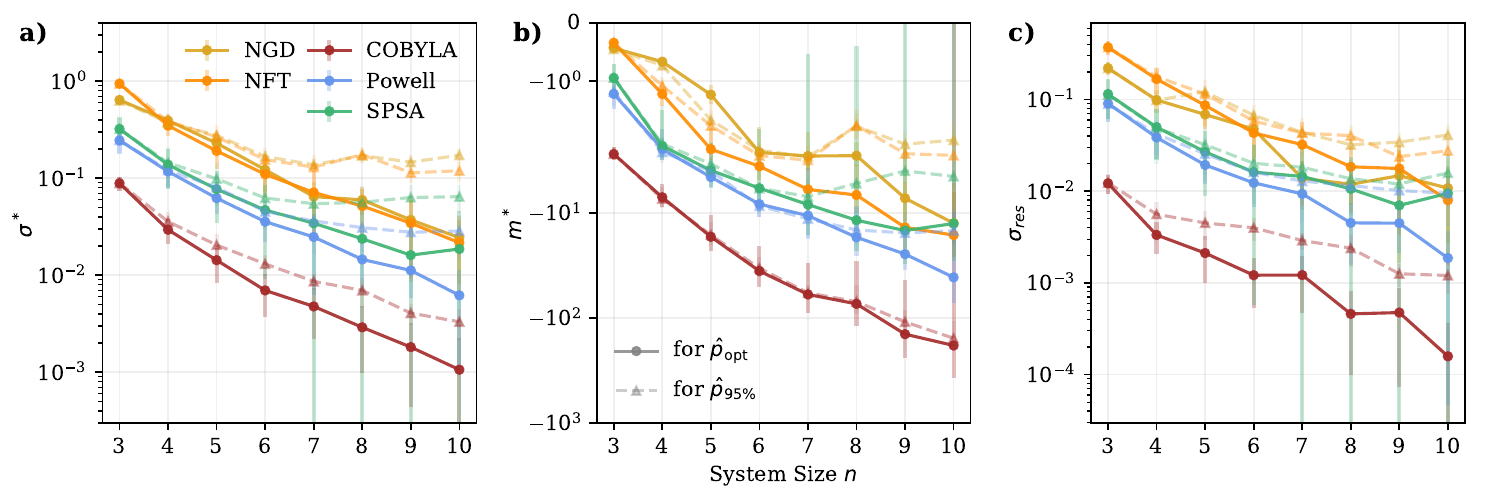}
\caption{\label{fig:scalingsystemsize} Noise resilience parameters as a function of system size. (a) The point of steepest descent $\sigma^*$, (b) slope at steepest descent $m^*$, and (c) point of resilience $\sigma_{\mathrm{res}}$ are shown for increasing system sizes $n$, derived from the $\tanh$-fits of $\hat{p}_{\mathrm{opt}}(\sigma)$ (solid line) and $\hat{p}_{95\%}(\sigma)$ (dashed line), for exact and approximate solvability of various optimizers, respectively. All plots use logarithmic y-axes for better visualization.}
\end{figure*}

To investigate the transitions dividing solvable and unsolvable regions in \cref{fig:gridplot}, the noise resilience parameters derived in the previous section are used. The point of steepest descent $\sigma^*$ \eqref{eq:sigmastar}, characterizes the critical threshold beyond which optimization fails. \cref{fig:scalingsystemsize} illustrates the behavior of this point, alongside its slope $m^*$ \eqref{eq:mstar}, and the point of resilience $\sigma_{\mathrm{res}}$ \eqref{eq:sigmares} across different system sizes. Together, these parameters quantify the practical limitations of each optimizer, providing a clear ranking of their ability to handle noise: NGD and NFT perform best, followed by SPSA, Powell, and finally COBYLA. These findings align well with related work~\cite{Nakanishi.2019,Singh.2023,PellowJarman.2021,Palackal.2023}. For a more detailed ranking and further discussion, refer to \cref{appsec:optimizerranking}.

Beyond ranking the robustness of the optimizers, we examine the overall scaling behavior of noise resilience metrics, which appear to follow consistent trends across different optimizers. Due to the strong similarity in results between $\hat{p}_{\mathrm{opt}}$ and $\hat{p}_{99\%}$, \cref{fig:scalingsystemsize} focuses on exact solvability ($\hat{p}_{\mathrm{opt}}$) and $\hat{p}_{95\%}$ for approximate solvability. 

The point of steepest descent $\sigma^*$ and the point of resilience $\sigma_{\mathrm{res}}$ should, in principle, exhibit similar scaling behavior, as both correspond to specific positions along the same $\tanh$-curve. All three metrics, including the slope of the descent's tangent line $m^*$, show a notable decline with increasing system size. While this overall trend is consistent across both exact and approximate solvability, the lower success threshold $\hat{p}_{95\%}$ intuitively exhibits a slightly softer decay compared to $\hat{p}_{\mathrm{opt}}$. However, since the practical usefulness of the approximate solvability metric is subject to limitations (see discussion in \cref{ssec:successmetric} and \cref{appsec:optimalitydistribution}), the subsequent analysis focuses on the scaling behavior of exact solvability. We emphasize, however, that the analysis was conducted for both thresholds, and full data is reported for each (see \cref{appsec:fitanalysis}). The overall qualitative conclusions remain unchanged, with only minor quantitative differences in the experimental curves.

\begin{figure}
\includegraphics[scale=0.69]{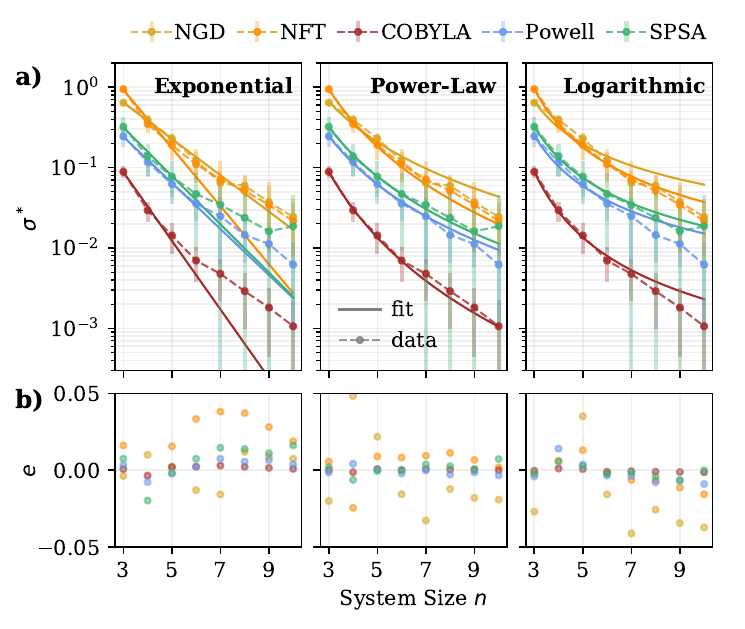}
\caption{\label{fig:fittingexperiments} Comparison of decay function fits. (a) Three theoretical decay functions (cf.~\cref{eq:fit_exp,eq:fit_pl,eq:fit_log}) are fitted to the experimental data, modeling the decline of noise resilience $\sigma^*$ with increasing system size $n$ of the classical optimizers. (b) The final fit residuals $e$ \eqref{eq:fitresiduals} show no significant differences among the three fit models, suggesting comparable accuracy in capturing the observed decay behavior.}
\end{figure}

Given the limited experimental data -- constrained by the computational cost of simulating larger system sizes -- any conclusions about the exact scaling behaviors or asymptotic trends require cautious analysis. To better understand the nature of the observable decay trends in \cref{fig:scalingsystemsize}, we therefore fit three different decay functions to the boundary line $\sigma^*(n)$:
\begin{flalign}\quad\sbullet[0.75]\quad \mathrm{exponential:}\quad
        f_{\exp}(n,k,\gamma)=k\:\exp({-}\gamma\:n) && \label{eq:fit_exp} \end{flalign}
\vspace{-2em}
\begin{flalign}\quad\sbullet[0.75]\quad \mathrm{power\text{-}law:}\quad
        f_{\mathrm{pl}}(n,k,\gamma)=k\: n^{-\gamma} && \label{eq:fit_pl} \end{flalign}
\vspace{-2em}
\begin{flalign}\quad\sbullet[0.75]\quad \mathrm{logarithmic:}\quad
        f_{\log}(n,k,\gamma)=k\: \log(n)^{-\gamma} && \label{eq:fit_log} \end{flalign}        
The results of these fits are illustrated in \cref{fig:fittingexperiments} for $\hat{p}_{\mathrm{opt}}$. Visually, the power-law fit appears to align best with the majority of optimizer curves. However, the fit residuals do not necessarily support this hypothesis. They are given by
\begin{equation}
e(n) = \sigma^*(n) - f(n, \hat{k}, \hat{\gamma}),
\label{eq:fitresiduals}
\end{equation}
where $e(n)$ represents the total difference between the measured $\sigma^*(n)$ values and the fitted decay function $f(n)$. Notably, none of the goodness-of-fit metrics were conclusive (cf.~\cref{appsec:fitanalysis}), indicating that the precise scaling behavior of noise resilience may vary across different optimizers. A larger data set with a broader range of optimizers would be required to confirm this hypothesis.

Despite uncertainties regarding the exact form of decay, one clear conclusion emerges: noise resilience declines rapidly across all tested methods. This decline is intuitively attributed to the corresponding shrinkage of loss function variance (cf.~\cref{eq:samplevariance}). However, even after normalizing the noise level $\sigma$ to account for this effect (cf.~\cref{sec:implications}), this decline persists. This suggests that loss variance decay cannot fully explain the observed drop in noise resilience among classical optimizers.

Several other factors must therefore contribute to this behavior: For instance, Fontana et al.~\cite{Fontana.2022} demonstrate that quantum noise can disrupt parameter symmetries, eliminating degeneracies in the loss landscape and complicating optimization. Additionally, Wang et al.~\cite{Wang.2021} identify the emergence of barren plateaus caused by noise, which severely impairs the trainability of VQAs. Other factors, such as the growing number of local minima in variational loss landscapes as system size grows~\cite{Rajakumar.2024}, and the issue of overparameterization~\cite{Kim.2022}, may also play a role. More broadly, Leymann and Barzen~\cite{Leymann.2020} provide a summary of the challenges associated with the trainability of VQAs, beyond the issue of barren plateaus.

For reproducibility, the final parameters, $k^*$ and $\gamma^*$, of all fitted theory curves (cf.~\cref{eq:fit_exp,eq:fit_pl,eq:fit_log}) are provided in \cref{appsec:fitanalysis}.

\section{\label{sec:implications}Practical Implications}

In recent years, numerous significant studies have highlighted the practical limitations of variational quantum algorithms~\cite{Akshay.2020,Bittel.2021,StilckFranca.2021,Fontana.2021,Wang.2021,Anschuetz.2022,GonzalezGarcia.2022,Scriva.2024}. However, the implications of these findings are often left to the reader or deferred to future research. In this section, we aim to bridge this gap by not only discussing the key insights from our results but also by explicitly demonstrating their practical relevance and broader impact.

Our analysis demonstrates that the noise resilience of VQA optimization decreases rapidly as the system size increases. This behavior was quantified in terms of the standard deviation of the total error on the loss function $\sigma^*$, beyond which the optimizer performance dropped to less than $50\%$ of its noise-free level. Among the tested optimizers, the NGD (normalized gradient descent) algorithm~\cite{Meli.2023} showed the best noise resilience.

To generalize these findings in this section, we use an error measure that is independent of the shrinking variance of the tested loss function: the Relative Absolute Error (RAE), defined as
\begin{equation}
    \epsilon(\mathcal{L}, \tilde{\mathcal{L}}) = \frac{\sum_{i=1}^N\ \lvert\tilde{\mathcal{L}}(\boldsymbol{\theta}_i)-\mathcal{L}(\boldsymbol{\theta}_i)\rvert}{\sum_{i=1}^N\ \lvert\mathcal{L}(\boldsymbol{\theta}_i)-\overline{\mathcal{L}}\rvert},
\label{eq:relativeabsoluteerror}
\end{equation}
where $\tilde{\mathcal{L}}$ denotes the noisy loss values, and $\overline{\mathcal{L}}$ is the sample mean. The RAE normalizes the mean absolute error by the mean absolute deviation of the loss values, making it a relative measure independent of the chosen normalization (cf.~\cref{eq:normalizedlossfunction}). 
Assuming an approximately normal distribution of the errors, the noise threshold, $\sigma^*(n)$, characterized in this work, can be translated into the RAE metric as:
\begin{equation}
    \epsilon^*(n) = \frac{\sigma^*(n)}{\sqrt{\mathrm{Var}(n)}},
\label{eq:sigma2rae}
\end{equation}
where $\mathrm{Var}(n)$ is the sample variance of the loss function, as depicted in \cref{fig:varianceanalysis}. 

\subsection{\label{ssec:imp_hardwaredevelopment}Hardware Development}

Despite significant advancements in quantum hardware development, current devices still face persistent limitations~\cite{Ezratty.2023}, with no demonstration of practical quantum advantage achieved to date. Mapping the RAE of loss functions when evaluated on current devices to the noise thresholds identified in this work provides valuable insights into these challenges. 

\begin{figure}
    \includegraphics[scale=0.69]{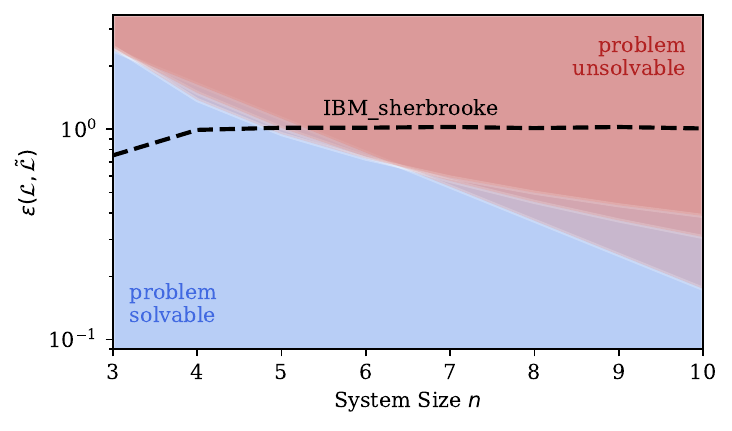}
    \caption{Exemplary hardware error and problem solvability regions. The three possible separation lines represent the three fitted functions (cf.~\cref{fig:fittingexperiments}) of the critical noise threshold $\sigma^*$ for the most noise-resilient optimizer, NGD, converted into the RAE metric $\epsilon^*$ \eqref{eq:sigma2rae} based on measurements from \cref{ssec:2lossfunctions}. Above these line, the probability of finding the optimal solution becomes negligible or effectively zero. The dashed line shows the empirically measured hardware error obtained using the \texttt{ibm\_sherbrooke}\footnote{Device specifications, based on the Eagle processor family: \num{7.095e-3} (\texttt{ECR} error), \num{2.184e-4} (\texttt{SX} error), \num{1.270e-3} (readout error), \SI{280.82}{\mu\s} ($t_1$), and \SI{204.19}{\mu\s} ($t_2$).} backend model.}
    \label{fig:3_2imphardware}
\end{figure}

\cref{fig:3_2imphardware} illustrates where the error produced by a representative IBM hardware noise model lies relative to the resilience thresholds of the most resilience optimizer, NGD (cf.~\cref{fig:fittingexperiments,appsec:fitanalysis}). The intersection of the two curves indicates that problems with more than six nodes may already be unsolvable on current hardware, according to our metric of solvability (cf.~\cref{eq:optimalsolutionprobability}).

While other qubit technologies or newer hardware may exhibit lower noise levels for the algorithm under consideration, the implications of the near-exponential scaling of precision requirements are clear: drastic reductions in hardware error rates are required to achieve practical applicability. This critical relationship between system size and noise levels has also been highlighted in previous studies~\cite{GonzalezGarcia.2022,StilckFranca.2021}.

According to IBM's Development \& Innovation Roadmap~\cite{IBMQuantum.2024}, the Starling processor, set for release in 2029, is expected to be error-corrected with 200 logical qubits. Similarly, Quantinuum's roadmap projects a nearly fault-tolerant quantum computer with 1000 logical qubits by 2030~\cite{Quantinuum.2024}. Comparable goals have been outlined by companies such as IonQ, IQM, and Google. This raises another critical question: will fully fault-tolerant hardware be able to address the scaling challenges and noise constraints highlighted in this work? In the following section, we seek to address this question.

\subsection{\label{ssec:imp_finitesamplingcurse}The Curse of Finite Sampling}

In the case of fault-tolerant hardware, the only persistent source of stochastic noise is produced by finite measurement sampling. Independent of whether VQAs will be used with fault-tolerant systems, this can be considered the ideal setting for these algorithms. This section, therefore, deals with the question of how the noise thresholds for problem solvability identified in \cref{sec:results} translate to the required number of measurement shots and what implications their scaling behavior has for achieving a practical quantum speed-up. In the following, we consolidate all previous findings, adopting the most optimistic scenarios.

Based on the results of \cref{ssec:scalingsystemsize}, the decline in an optimizer's noise resilience as the system size $n$ increases, denoted by $\sigma^*(n)$, may follow an exponential, logarithmic, or power-law trend (cf.~\cref{fig:fittingexperiments}). To convert these resilience thresholds of the most noise-resilient optimizer, NGD, into the general measure of the RAE \eqref{eq:relativeabsoluteerror}, they need to be normalized by the variance of the corresponding loss function (cf.~\cref{eq:sigma2rae}), as obtained for BENQO. 
The most optimistic scaling for the loss variance, which also provided a good fit, followed a power-law decay \footnote{Using an exponential fit for the loss variance -- consistent with a barren plateau assumption -- would imply an increase in the noise resilience for system sizes $n\geq61$ in the power-law case \eqref{eq:fit_pl}, and $n\geq27$ for logarithmic decay \eqref{eq:fit_log}. While this would not change the overall implications of \cref{ssec:imp_finitesamplingcurse}, this scenario is unlikely.}. Using the fit parameters from \cref{fig:varianceanalysis}, the maximum RAE $\epsilon^*(n)$ that an optimizer can tolerate for a given system size $n$ computes to
\begin{equation}
    \epsilon^*(n) \approx
    \begin{dcases*}
        8.5\: n^{0.4}\: \mathrm{e}^{-0.5\:n}
       & ${\scriptstyle \mathrm{for\ exponential}}$\\
        22.0\: n^{-1.8}
       & ${\scriptstyle \mathrm{for\ power\text{-}law}}$\\
        2.5\: n^{0.4}\: \log(n)^{-3.2}
       & ${\scriptstyle \mathrm{for\ logarithmic}}$
    \end{dcases*},
\label{eq:raeboundscaling}
\end{equation}
depending on the realized scaling law. These results, while based on numerical fits with inherent uncertainties, strongly suggest that the decline in resilience persists regardless of how the empirical data is fitted \footnote{If $\sigma^*$ were to decay logarithmically, $\epsilon^*$ would eventually start increasing again due to the faster power-law decay of the loss variance. However, since this effect only becomes relevant for ${n > \mathrm{e}^8\approx2981}$, it remains negligible within the scope of our findings.}.

\begin{figure}
\includegraphics[scale=0.69]{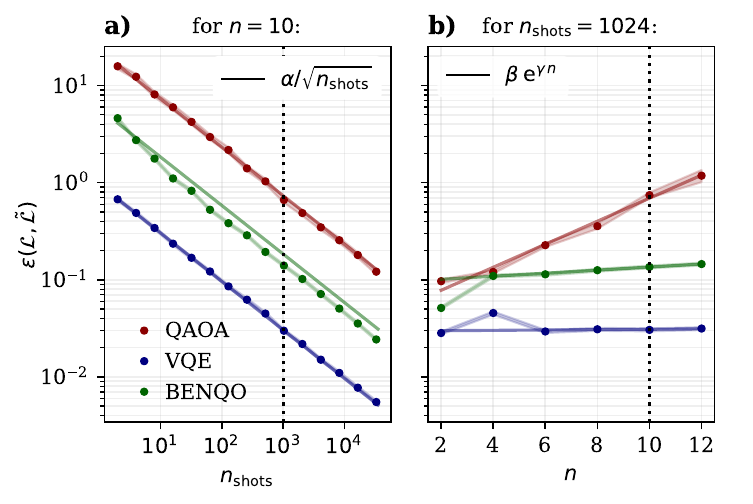}
\caption{\label{fig:FSerrors} Relative absolute error $\epsilon$ induced by finite sampling of the loss function. (a) The RAE \eqref{eq:relativeabsoluteerror} is plotted as a function of measurement shots $n_{\mathrm{shots}}$ (logarithmic scale) for a fixed system size $n=10$, along with corresponding $1/\sqrt{n_{\mathrm{shots}}}$-fit functions. (b) The RAE is shown for increasing system sizes $n$ at a fixed $n_{\mathrm{shots}}=1024$ shots, a common default in experimental computations, with exponential fit functions applied for $n\geq6$ to reduce small-scale effects. Vertical dotted lines indicate the region of overlap between both plots. All error values are computed from 1000 loss evaluations per $n$, with parameters uniformly sampled from $[-2\pi,2\pi]^n$.}
\end{figure}

To map this to measurement shot requirements, we first analyze the error introduced by finite sampling in evaluating a loss function, denoted as $\epsilon_{\mathrm{FS}}$. \cref{fig:FSerrors} illustrates the behavior of the RAE \eqref{eq:relativeabsoluteerror}, computed for three VQAs (cf.~\cref{ssec:2lossfunctions}), including BENQO, as a function of the number of shots $n_{\mathrm{shots}}$ (at a fixed system size) and the system size $n$ (at a fixed number of shots). As expected from a multinomial probability distribution, the error decreases proportionally to $1/\sqrt{n_{\mathrm{shots}}}$ when $n$ is fixed. Conversely, when $n_{\mathrm{shots}}$ is kept constant, the error grows exponentially, reflecting the increasing range of loss values alongside a simultaneously shrinking variance as the system size increases. For an analytical derivation of this scaling behavior, see \cref{appsec:FSerrordistribution} (cf.~\cite{Barligea.2024}).

Assuming that this general scaling behavior is consistent across all sampling rates and system sizes, we derive an approximate function for the RAE due to finite sampling for the examined loss function. Using the measured error value at $(n=10, n_{\mathrm{shots}}=1024)$, where both experimental curves align, and applying the fitted parameters for BENQO from \cref{fig:FSerrors}, the finite sampling error for system size $n$ and $n_{\mathrm{shots}}$ measurement shots can be expressed as:
\begin{equation}
   \epsilon_{\mathrm{FS}}(n, n_{\mathrm{shots}})\approx 4.0\: \frac{\mathrm{e}^{0.04\:n}}{\sqrt{n_{\mathrm{shots}}}}.
\label{eq:raeofFS}
\end{equation}

For a random QUBO instance to be solvable with a VQA -- specifically BENQO -- the solvability condition 
\begin{equation}
\epsilon_{\mathrm{FS}}(n, n_{\mathrm{shots}})< \epsilon^*(n)
\label{eq:criterionofsolvability}
\end{equation}
must hold for all $n$. Substituting \cref{eq:raeboundscaling} and \cref{eq:raeofFS}, we derive that the required number of measurement shots grows exponentially with system size, for the scaling assumed in \cref{eq:raeofFS}. Specifically, 
\begin{equation}
n_{\mathrm{shots}}(n) > \left( \frac{4.0\:\mathrm{e}^{0.04\: n}}{\epsilon^*(n)}\right)^2
\label{eq:3_2scalingofrequiredshots}
\end{equation}
is always of order $\mathcal{O}(k^n)$ for some $k>1$, regardless of the specific scaling law in \cref{eq:raeboundscaling}. This aligns with previous studies reporting exponential resource requirements in variational quantum optimization~\cite{Mazzola.2022,Scriva.2024,Wang.2024}.

Furthermore, Scriva et al.~\cite{Scriva.2024} emphasized that the quantum computational cost, quantified as $n_{\mathrm{shots}}\cdot n_{\mathrm{calls}}$, must strictly remain below $2^n$ to avoid quantum disadvantage -- i.e., the point where classical brute-force algorithms become more efficient than quantum computing. Note that this is the case even without accounting for quantum gate execution times. 

\begin{figure}
\includegraphics[scale=0.69]{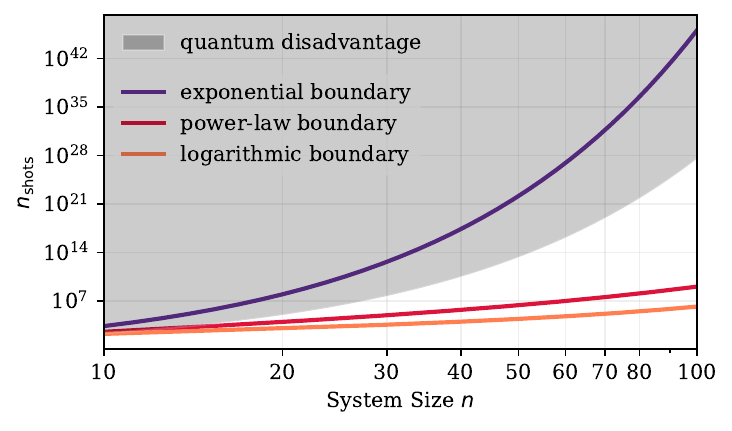}
\caption{\label{fig:imp_shotscaling} Required shots $n_{\mathrm{shots}}$ for solving a random QUBO problem of size $n$. The minimum number of shots required by the most noise resilient optimizer, NGD, to reach three specified noise thresholds is shown as solid lines. For any value above these theoretical curves, solving a QUBO problem of size $n$ using the BENQO algorithm becomes feasible. The shaded region represents the regime where the number of required quantum circuit executions surpasses the number of classical brute-force trials, rendering the quantum algorithm less efficient than the worst classical solver.}
\end{figure}

In our experiments, the most noise-resilient optimizer, NGD, requires only 20 iterations regardless of $n$, which each consist of one quantum call to evaluate the loss function and $2n$ additional calls to evaluate the gradient (cf.~\cref{eq:parametershiftrule}). For this optimizer, the maximum number of allowed shots before encountering quantum disadvantage is given by:
\begin{equation}
n_{\mathrm{shots}}(n) < \frac{2^n}{20(1+2n)}.
\label{eq:shotsforquantumdisadvantage}
\end{equation}
This implies that for problem sizes below $9$, achieving a quantum speed-up is impossible, as brute-force solutions are always more efficient. \cref{fig:imp_shotscaling} graphically summarizes these findings for system sizes $n\geq10$.

The figure clearly illustrates that if the boundary line identified in \cref{ssec:scalingsystemsize} was to follow an exponential decay, the number of required measurement shots would already grow faster than the number of brute-force trials needed to solve the problem classically. This strongly suggests that quantum advantage for optimization tasks under these conditions is fundamentally unattainable. In the alternative scenarios of power-law and logarithmic scaling, a small window of potential quantum advantage remains for problem sizes $n\geq20$ (logarithmic boundary) and $n\geq25$ (power-law boundary), where the shot requirement still falls below the critical threshold of quantum disadvantage. However, these potential advantages are only theoretical, as they are based purely on resource scaling and do not yet translate into practical runtime improvements.

Therefore, we extend this analysis to the total runtime $t$ required for the optimization process. Since this time varies significantly across hardware platforms, Scriva et al.~\cite{Scriva.2024} proposed a general lower bound for the runtime:
\begin{equation}
t_{\mathrm{run}}<n_{\mathrm{shots}}\cdot n_{\mathrm{calls}}\cdot D\cdot t_{\mathrm{gate}}.
\label{eq:lowerboundtime}
\end{equation}
Here, $n_{\mathrm{calls}}$ denotes the number of quantum loss function evaluations, each measured with $n_{\mathrm{shots}}$. The circuit depth of the used ansatz is denoted by $D$, and $t_{\mathrm{gate}}$ is the hardware-specific time required to execute a single gate. Notably, $n_{\mathrm{calls}}$ does not necessarily correspond to the number of optimization iterations, as many optimizers require additional function evaluations per step, particularly for gradient estimation (e.g., $2n$ calls using the parameter-shift rule \eqref{eq:parametershiftrule}).

For problem sizes of practical relevance, ${n\geq\mathcal{O}(100)}$, even under the most optimistic assumptions -- constant circuit depth ${D=\mathcal{O}(1)}$ and gate execution times of approximately $100\: \text{ns}$ -- the lower bound on the time to solution would already exceed one hour. Given the fast scaling of this time and the fact that this comparison is only made against the worst classical algorithms (while state-of-the-art classical solvers can handle problem instances orders of magnitude beyond 100~\cite{Concorde,Applegate.2007,Pawlowski.2025} in less time), any potential quantum speed-up appears highly impractical. 

These findings further highlight the substantial gap that quantum devices still have to bridge to outperform classical solvers and achieve real world impact in optimization. They also cast significant doubt on the practical feasibility of achieving meaningful quantum advantage in this domain using variational quantum algorithms.

\section{\label{sec:validation}Validation and Discussion}

A common concern in empirical studies like ours is to what degree the observed results are influenced by specific choices in the experimental setup (cf.~\cref{fig:methodology}). Factors such as problem formulation, quantum ansatz, noise model, optimizer selection, and performance metrics may, in principle, affect the conclusions drawn. One could, therefore, argue that the critical behavior identified in this work merely arises from unlucky experimental choices rather than inherent scalability challenges of hybrid variational quantum optimization. While each of our decisions was carefully justified to favor well-behaved scenarios and ensure a fair and optimistic evaluation, this section provides additional validation and discussion about the reliability and generality of our findings.

\subsection{Independent Confirmation of Scaling Trends}
A study by Scriva et al.~\cite{Scriva.2024} examined shot noise effects using a VQE ansatz on an Ising chain problem and found that the success probability for a fixed system size follows the same sigmoidal pattern as observed in our experiments. Their results also demonstrated that the optimal number of function calls required to reach a fixed success probability scales exponentially with system size for certain parameter initializations, reinforcing our findings in \cref{ssec:imp_finitesamplingcurse}.

Similarly, Sung et al.~\cite{Sung.2020} analyzed a QAOA ansatz applied to three distinct optimization problems, including the MaxCut problem on three regular graphs. Unlike our study, they employed global and alternative classical optimizers and modeled noise at the gate level of the quantum ansatz. Regardless of these differences in setup, they also observed a sigmoidal drop in success probability as noise increased and an unfavorable scaling of the required time to solution for fixed precision levels.

Despite the differences in quantum algorithms, noise models, classical optimizer selection, and problem types, both studies independently observed the same key trends as those presented in this work. 

\subsection{Validation with a VQE Ansatz}
\begin{figure}
\includegraphics[scale=0.59]{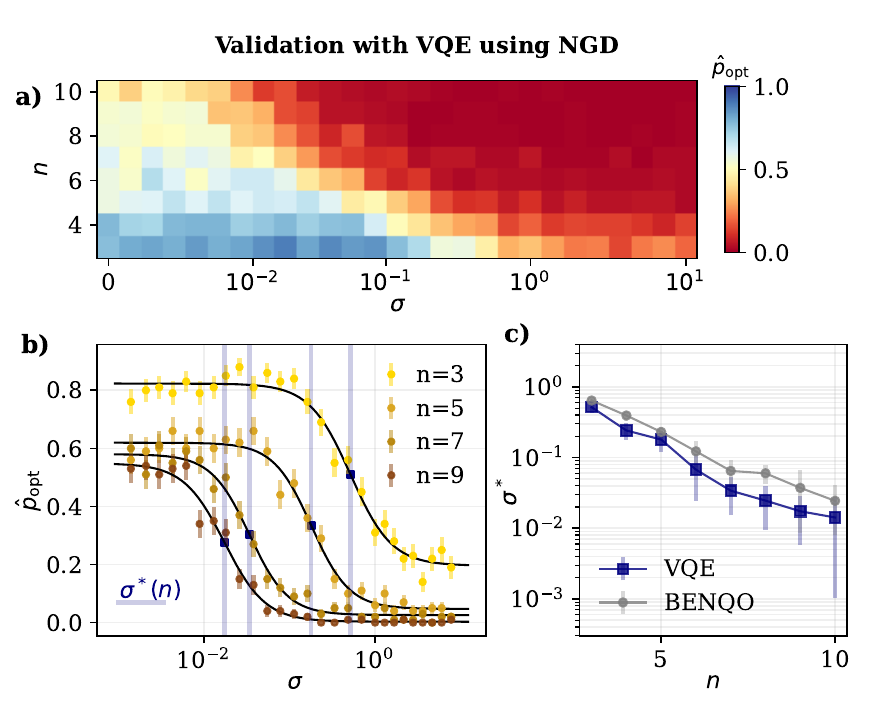}
\caption{\label{fig:val_VQEwithnoise_NGD}Validation of NGD solvability scaling with VQE. (a) Contour plot of the empirical NGD solvability experiments (cf.~\cref{fig:gridplot}), this time using VQE as the underlying quantum loss function. (b) Scaling of the success probability $\hat{p}_{\mathrm{opt}}$ across different noise levels for fixed system sizes, fitted with sigmoidal theory curves as in \cref{fig:scalingnoiselevel}. Blue vertical lines indicate the points of steepest descent $\sigma^*$ \eqref{eq:sigmastar}, obtained from $tanh$-fits (cf.~\cref{eq:tanhfitfunction}). (c) The main noise resilience parameter $\sigma^*$ is plotted against increasing system sizes and directly compared to the noise threshold obtained for BENQO (cf.~\cref{ssec:scalingsystemsize}). The curves exhibit a close agreement.}
\end{figure}

We repeat our experiments to further validate our conclusions using another well-behaved quantum loss function: the two-local shallow VQE ansatz described in \cref{ssec:2lossfunctions}. The results for the NGD optimizer are summarized in \cref{fig:val_VQEwithnoise_NGD}, closely matching those obtained using BENQOs. This agreement is not only qualitative but also quantitative. Normalizing the noise resilience threshold $\sigma^*(n)$ by the variance of the respective loss functions (cf.~\cref{fig:varianceanalysis}) -- thus mapping both results onto the RAE metric \eqref{eq:relativeabsoluteerror} -- would likely reveal an even stronger correspondence between the two ansatzes. 

To directly validate the corresponding resource scaling reported in \cref{ssec:imp_finitesamplingcurse}, we conduct additional experiments with the VQE algorithm, this time introducing noise solely through finite sampling rather than the additive Gaussian noise model. The shallowness of the VQE ansatz enables simulations with larger system sizes than in previous experiments, extending the range to $n\in[3,13]$. The following results focus on the COBYLA solver, as it demonstrated the highest resource efficiency in terms of quantum function calls and mean runtime (cf.~\cref{appsec:optimizerranking}). 

\begin{figure*}
\includegraphics[width=1\linewidth]{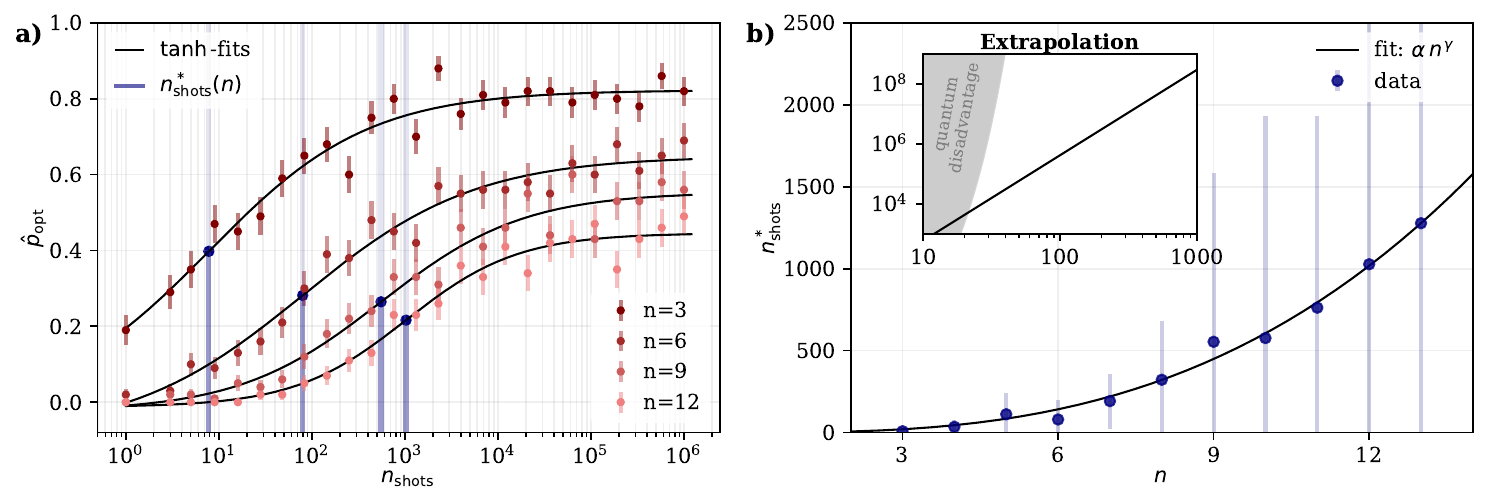}
\caption{\label{fig:val_VQEwithshotnoise_COBYLA}Validation of measurement shot scaling with VQE. (a) Success probability $\hat{p}_{\mathrm{opt}}$ as a function of the used numbers of measurement shots $n_{\mathrm{shots}}$ in simulation, for fixed system sizes $n$. Blue vertical lines mark the positions of steepest descent $n_{\mathrm{shots}}^*$, obtained from $tanh$-fits (cf.~\cref{eq:tanhfitfunction,eq:sigmastar}). (b) Required number of measurement shots for system solvability as a function of system size. A power-law fit of the data enables extrapolation to larger system sizes, as shown in the inset, with the region of quantum disadvantage shaded in gray (cf.~\cref{fig:imp_shotscaling}).}
\end{figure*}

\cref{fig:val_VQEwithshotnoise_COBYLA} presents the results of this analysis. As expected, we observe a sigmoidal drop in success probability for fixed system sizes $n$ as the number of measurement shots $n_{\mathrm{shots}}$ decreases. By fitting the data using \cref{eq:tanhfitfunction}, this time with $n_{\mathrm{shots}}$ as the variable, we extract the steepest descent positions $n_{\mathrm{shots}}^*$ (cf.~\cref{eq:sigmastar}) and analyze their scaling behavior with increasing system sizes. A power-law function provides the best visual agreement and is therefore used to extrapolate the required shot numbers to practically relevant system sizes. This allows for a direct comparison with \cref{fig:imp_shotscaling}. As before, the boundary of quantum disadvantage is derived by dividing the number of brute-force trials $2^n$ by the number of function calls (cf.~ \cref{eq:shotsforquantumdisadvantage}). 

These validation results align more closely with the power-law scaling scenario discussed in \cref{ssec:imp_finitesamplingcurse} and a power-law scaling of the finite sampling error $\epsilon_{\mathrm{FS}}$ \eqref{eq:raeofFS} in $n$, meaning that there is a theoretical window of quantum advantage -- though only when compared to the most inefficient classical solution method. Interestingly, our direct shot requirement analysis suggests slightly lower resource demands than initially estimated. However, under the same optimistic assumptions, i.e., a constant circuit depth ${D=1}$ and gate execution times of approximately $100\: \text{ns}$, the lower bound for time to solution \eqref{eq:lowerboundtime} would still exceed one hour for problems of size $n\geq243$. While this is slightly above the $n=100$ estimate from \cref{ssec:imp_finitesamplingcurse}, the difference remains minor given that these calculations assume best-case scenarios with only finite sampling errors -- an assumption that is only valid in fully error-corrected quantum devices.

\subsection{Reconsidering the Success Metric}
\label{ssec:successmetric}

Another point of discussion of our study is the choice of success metric. In this study, we measured the sample success probability of finding a solution within a specific relative threshold $t$ of the optimal energy (cf.~\cref{eq:optimalsolutionprobability}), with thresholds set at $t=100\%$ (exact optimality), $99\%$, and $95\%$ (approximate solvability). However, most of the scalability issues discussed are derived from the probability of finding the optimal solution $\hat{p}_{\mathrm{opt}}$. One could argue that the primary goal of quantum optimization, particularly for NP-hard problems, is not necessarily to find the absolute global optimum but rather to generate high-quality approximations that outperform classical solutions. 

Our experimental results indicate that the scaling behavior of $\hat{p}_{99\%}$ and $\hat{p}_{95\%}$ differs slightly from that of the exact solution probability (cf.~\cref{fig:scalingsystemsize}). In particular, the further the relative energy threshold $t$ deviates from $100\%$, the more its scaling diverges from the solvability scaling discussed earlier. In some cases -- especially in the absence of noise -- the near-optimal success probability $\hat{p}_{95\%}$ even increases with system size (cf.~\cref{fig:gridplot}), suggesting that as the system grows, optimizers tend to converge to local minima that lie within a close vicinity of the global optimum.

This trend can be explained by examining the distribution of loss values across all possible solution states (cf.~\cref{appsec:optimalitydistribution}). In our QUBO formulation, feasible solutions are enforced via quadratic penalty terms (cf.~\cref{appsec:tspconstraintsexample}), which can -- depending on the chosen penalty factor -- significantly expand the loss value range in one direction, often leading to $|C_{\max}|\gg|C_{\min}|$. Due to the chosen normalization, the maximum loss value remains 1, while the majority of loss values concentrate (possibly exponentially) near zero (cf.~\cref{ssec:2lossfunctions}), leading to an asymmetry in the loss value distribution. As the system size increases, this effect intensifies, effectively pushing most local minima closer to the global optimum. Consequently, the approximation ratio in \cref{eq:approximationindex} approaches 1, shifting the threshold for identifying feasible solutions to much higher energy levels -- sometimes even beyond the $99\%$ threshold used in this study~\cite{Salehi.2022,Palackal.2023}. This explains why approximate solutions remain relatively accessible, even when finding the exact optimum becomes increasingly difficult. 

However, this also raises doubts about the practical utility of near-optimal solutions using standard metrics as in \cref{eq:approximationindex}, particularly if they fail to satisfy feasibility constraints. One approach to mitigate this issue is to explore alternative problem encodings that yield a higher proportion of feasible solutions~\cite{Schnaus.2024}. Another possibility it to redefine the chosen success probability metric (cf.~\cref{eq:optimalsolutionprobability}). For this, rather than relying on the approximation ratio as defined in \cref{eq:approximationratio}, a more informative metric could be a normalized approximation ratio that considers the energy of a totally mixed state as the upper bound $C_{\max}$. This adjustment would provide a more meaningful indicator of high-quality solutions, rather than merely achieving proximity to the optimal value under a potentially misleading scale.

\subsection{On Ansatz Choice and Smart Initializations}
\label{ssec:ansatzchoiceinitializations}
In this study, we deliberately focused on two variational ansatzes that avoid barren plateaus, maintaining non-vanishing gradient and loss values to ensure more favorable trainability properties on average (cf.~\cref{ssec:2lossfunctions}). However, their absence of barren plateaus is ultimately linked to a non-exponentially scaling dimension of the underlying dynamical Lie algebra~\cite{Ragone.2024,Fontana.2024}, which in turn implies efficient classical simulability~\cite{Cerezo.2023,Goh.2025}. As such, these ansatzes may not offer a viable path toward achieving quantum advantage within VQAs. 

To address this, one might intentionally consider ansatz structures that do exhibit BPs, such as QAOA or generic hardware-efficient ansatzs (HEAs)~\cite{Larocca.2022}, in order to retain the possibility of quantum advantage in the loss computation. In these cases, the key challenge becomes overcoming the exponential suppression of loss and gradient magnitudes. A leading approach to mitigating this issue is the use of smart parameter initialization strategies, which have shown promising performance in both QAOA~\cite{Egger.2021,Sack.2021,Scriva.2024,Boulebnane.2021,Jain.2022,NCliz.2025} and HEA circuits~\cite{Grant.2019,Wang.2024b,Zhang.2022b,Volkoff.2021,Park.2024}. An important direction for future work is to investigate how the presence of Gaussian uncertainty interacts with such initialization strategies, and whether the observed impact of stochastic noise differs when applied to barren-plateau-prone ansatzes initialized with trainability-enhancing methods.

\section{\label{sec:conclusions}Conclusions and Outlook}

The potential of variational quantum algorithms for optimization fundamentally depends on their ability to handle stochastic noise, which arises both from current hardware imperfections and finite sampling -- the latter remaining even in fault-tolerant quantum computers. While numerous theoretical and empirical studies have highlighted how hardware noise can impair or even eliminate quantum advantage in VQAs (cf.~\cref{sec:relatedwork}), there has been less focus on the impact of isolated stochastic uncertainty on system solvability. This work provides a rigorous numerical investigation into how classical optimizers -- the main computational engine of VQAs -- respond to additive Gaussian noise as the size of quantum loss functions scales up. Our findings, validated through extensive testing, empirically demonstrated that stochastic noise alone may impose fundamental limitations on the scalability of variational quantum optimization to practically relevant problem sizes, even when using fault-tolerant hardware.

We find, that for fixed system sizes, the probability of finding high-quality solutions -- including those within $1\%$ and $5\%$ of the optimal energy -- decreases in a sigmoidal pattern with respect to the logarithmic noise level. This reveals a sharp performance drop at a critical noise threshold $\sigma^*$, which varies significantly across the different optimizers. Using $\tanh$-fits, we extracted these thresholds for system sizes $n\in[3,10]$, suggesting a common scaling trend across all optimizers: their noise resilience declines rapidly, potentially exponentially, as system size increases. Notably, this decline is steeper than what can be explained by a simultaneously shrinking loss variance, confirming that larger systems inherently present a more challenging loss landscape.

While it is well known that current hardware error rates far exceed what is required for classically relevant problem sizes, our results highlight another possible limitation imposed by finite sampling errors. Extrapolating the observed decline in noise resilience showed that, to achieve the loss precision necessary for problem solvability, the required number of measurement shots scales prohibitively bad with system size. This leaves minimal room before exceeding the computational resources of classical brute-force sampling, which is highly impractical. These findings suggest that achieving meaningful quantum advantage with VQAs for large-scale optimization problems may face significant challenges.

We therefore advocate for a shift in research focus -- away from introducing yet new heuristic hybrid algorithms with idealized testing, and toward strategies that directly address the challenges demonstrated in this and prior work. If quantum utility~\cite{Herrmann.2023} in optimization remains the goal, future research should prioritize alternative algorithmic frameworks (and possibly different problem domains) that can provably circumvent these barriers.

Apart from the methodological suggestions made in \cref{ssec:successmetric,ssec:ansatzchoiceinitializations}, one promising avenue is improved problem encoding, which aims to reduce the number of required qubits and thereby mitigate some of the scaling issues discussed here~\cite{Sciorilli.2024,Bermejo.2023,Maciejewski.2024,Larsen.2024,Jiang.2025}. However, while these methods can offer potential efficiency gains, they often lack both theoretical performance guarantees and large-scale empirical validation in noise-exposed settings. Moreover, given that the scalability challenges demonstrated here may be inherent in the hybrid nature of VQAs, which rely on classical optimization, it seems to be more promising to explore fully quantum alternatives outside the variational paradigm. Recent works~\cite{Bennett.2024,Catli.2025,Jordan.2025,Jiang.2025} have begun such investigations, and a similar shift away from hybrid frameworks can be seen in the quantum machine learning community~\cite{Huang.2024}. Additionally, this study has not yet considered alternative computational models, such as Quantum Boson Sampling~\cite{Hamilton.2017,Slysz.2024,Go.2025} or Quantum Annealing~\cite{Finnila.1994,Yarkoni.2022}, which may warrant further investigation.

Beyond algorithmic strategies, a re-evaluation of the types of problems targeted for near-term quantum advantage may be necessary. Rather than tackling intrinsically classical problems such as Ising spin glasses, growing attention is being given to quantum-native optimization problems, such as the simulation of quantum mechanical systems~\cite{Feynman.1982,Lloyd.1996}, where exponential speed-ups are both theoretically proven and more intuitive. For recent advances in quantum simulation, see Refs.~\cite{Daley.2022,Fauseweh.2024}.

Ultimately, our work provides a general and robust framework for rigorously investigating the scalability and limitations of not only VQAs but also other quantum algorithms in the presence of stochastic errors. We encourage researchers to apply and extend our methodology to more promising alternative setups -- whether by testing different quantum ansatzes, optimizer setups, problem instances, or hardware platforms. If future studies uncover scenarios where problem solvability scales more favorably with system size and noise levels, this would be an exciting breakthrough. However, if the scaling behavior observed here proves to be universal and inherent to the classical optimization problem inherent in hybrid architectures like VQAs, it would strongly suggest that quantum advantage in hybrid variational quantum optimization is not realistic. In fact, our results indicate that the challenges facing VQAs -- both in the NISQ and fault-tolerant era -- may be more fundamental than what can be resolved through hardware improvements, better classical optimizers, or ansatz tuning alone.

\begin{acknowledgments}
This work is supported by the Federal Ministry for Economic Affairs and Climate Action on the basis of a decision by the German Bundestag through
the project \textit{QuaST -- Quantum-enabling Services and Tools for Industrial Application}. 
The authors would like to express their gratitude to Frank Pollmann for providing valuable advice on this work and carefully reviewing this manuscript.
We also thank Marita Oliv, Amine Ben, and Nicola Franco for insightful discussions and valuable feedback, which have contributed to the refinement of this study.
\end{acknowledgments}

\appendix
\section{\label{appsec:qubo2isingmapping}Mapping QUBO to Ising Problems}

To translate the general QUBO form \eqref{eq:QUBOproblem}
\begin{equation}
    f(x)=\sum_{i=1}^{n} \sum_{j=1}^{n} Q_{i j}\: x_{i}\: x_{j}
\end{equation}
into the Ising form of \cref{eq:IsingCostFunction}, one can apply the mapping ${x_i=\frac{z_i+1}{2}}$ between the binary variables ${x_i\in\{0,1\}}$ and the spin variables ${z_i\in \{-1,+1\}}$ as follows:
\begin{equation}
    \begin{aligned}g(z) & =\sum_{i=1}^{n} \sum_{j=1}^{n} Q_{i j}\left(\frac{z_{i}+1}{2}\right)\left(\frac{z_{j}+1}{2}\right)\\
&= \sum_{i=1}^{n} \sum_{j=1}^{n} \biggl(\frac{Q_{i j}}{4} z_{i} z_{j}+\frac{Q_{i j}}{4} z_{i} +\frac{Q_{i j}}{4} z_{j}+ \underbrace{\frac{Q_{i j}}{4}}_{\text{const.}} \biggr) \\& \sim\sum_{i\neq j} \frac{Q_{i j}}{4} z_{i} z_{j}
+\sum_{i=1}^{n}\left(\sum_{j=1}^{n} \frac{Q_{i j}}{4}+\sum_{j=1}^{n} \frac{Q_{j i}}{4}\right) z_{i}
\end{aligned}
\end{equation}
In the final step, all constant offsets were disregarded as they do not affect the minimization of the loss function.

Assuming that $\mathbf{Q}$ is a symmetric matrix, and by replacing the spin variables $z$ with Pauli-$\mathbf{Z}$ operators, the final Ising operator depending on the QUBO matrix elements becomes:
\begin{equation}
\mathbf{C}(\mathbf{Q}) =\sum_{i<j} \frac{Q_{i j}}{2} \mathbf{Z}_{i} \mathbf{Z}_{j}+\sum_{i=1}^{n}\left(\sum_{j=1}^{n} \frac{Q_{i j}}{2}\right) \mathbf{Z}_{i}
\end{equation}
Generally, this reformulation means that the Ising weight matrix $\boldsymbol{\mathcal{C}}=\{\mathcal{C}_{ij}\}$ (cf.~\cref{eq:IsingCostFunction}) can be retrieved from the QUBO matrix $\mathbf{Q}=\{Q_{ij}\}$ as follows:
\begin{itemize}
\setlength\itemsep{0.1em}
    \item off-diagonal elements: $\mathcal{C}_{ij}=Q_{ij}/2$
    \item diagonal elements: $\mathcal{C}_{ii}=\sum_{j=1}^n Q_{ij}/2$
\end{itemize}
A similar mapping between the QUBO and Maximum Cut (MaxCut) problem was derived in Ref.~\cite{Barahona.1989}. 

Lastly, note that this transformation is not unique. An equivalent formulation can be obtained by applying the mapping $x_{i}=\frac{1-z_{i}}{2}$, which transforms the binary variables $x\in\{0,1\}$ into the spin variables $z\in\{+1,-1\}$. This leads to an Ising form with a negative sign for the diagonal terms. In this formulation, the ground state configuration would have $0$s where the original formulation had $1$s, and vice versa. Thus, this mapping would result in an inverted ground state compared to the initial transformation.

\section{\label{appsec:tspconstraintsexample}Quadratic Penalty Terms in Permutation Problems}

In graph partitioning problems, such as the well-known MaxCut problem, the number of nodes in the problem graph directly corresponds to the number of binary variables in the problem formulation. However, permutation problems, such as the highly relevant Traveling Salesperson Problem (TSP), are more complex as they require assigning an order instead of a partition to the nodes. This can be achieved using one-hot-encoding, where permutations are represented by $n^2$ binary decision variables $x_i^{\alpha}$, indicating whether node $i$ is assigned to a position $\alpha$ in the permutation. In the case of the TSP, $i$ would represent a specific city, and $\alpha$ denotes the time step at which the city is visited along the route. 

This binary encoding is therefore subject to two permutational constraints:
\begin{itemize}
\setlength\itemsep{0.1em}
    \item Each position $\alpha$ must hold exactly one node:\\$\sum_i x_i^{\alpha} = 1\quad \forall\ \alpha=1,\dots,n$
    \item Each node $i$ must occur exactly once:\\$\sum_{\alpha} x_i^{\alpha} = 1\quad \forall\ i = 1,\dots ,n$
\end{itemize}
Glover et al.~\cite{Glover.2022} proposed incorporating these constraints into a QUBO loss function like \cref{eq:QUBOproblem} using quadratic penalty terms with a penalty factor $P>1$, resulting in the new QUBO formulation:
\begin{equation}
    C(\mathbf{x}) = C_0(\mathbf{x}) + P\:\biggl[ \sum_{i=1}^n\bigl(1-\sum_{\alpha=1}^n x_i^\alpha\bigr)^2+\sum_{\alpha=1}^n\bigl(1-\sum_{i=1}^n x_i^\alpha\bigr)^2\biggr].
\label{eq:problemformulation}
\end{equation}
Here, $C_0(\mathbf{x})$ represents the actual unconstrained loss function of the problem, defined by the QUBO matrix $\mathbf{Q_0}$, while the latter terms impose the corresponding permutational constraints. In the TSP, $C_0(\mathbf{x})$ would be a function returning the total path length. 

To ensure that feasible solutions are enforced after minimizing \cref{eq:problemformulation}, $P$ must be larger than a problem-specific threshold dependent on $C_0$. 
To obtain the QUBO matrix representation from the loss function in \cref{eq:problemformulation}, we translate its terms into the matrix elements $Q_{i j}^{\alpha \beta}$, which encode the interaction between variables $x_{i\alpha}$ and $x_{j \beta}$:
\begin{equation}
 \begin{aligned}
Q_{i j}^{ \alpha \beta}=&\  \{\mathbf{Q_0}\}_{i j}^{ \alpha \beta} \\& 
+P\left[\delta_{i j}\left(1-\delta_{\alpha \beta}\right)+\delta_{\alpha \beta}\left(1-\delta_{i j}\right)\right]\\&
-4\:P\: \delta_{i j}\: \delta_{\alpha \beta}.
\label{eq:specialqubo}
\end{aligned}   
\end{equation}
From \cref{eq:specialqubo}, it is immediately apparent that the penalties terms affect the diagonal and off-diagonal elements of $\mathbf{Q}$ differently: 
\begin{itemize}[leftmargin=*]
\setlength\itemsep{0.1em}
    \item The off-diagonal penalty terms (second row of \eqref{eq:specialqubo}) introduce additional contributions of order $P$ to the original problem matrix $\mathbf{Q_0}$.
    \item The diagonal penalty terms (third row of \eqref{eq:specialqubo}) shift the diagonal elements in $\mathbf{Q_0}$ downward by values of order $4P$.
\end{itemize}
 Since $P$ is chosen to be significantly larger than each element of $\mathbf{Q_0}$, all diagonal elements of $\mathbf{Q}$ become negative. This confirms that adding equality constraints as quadratic penalty terms transforms the problem's QUBO matrix into one with negative diagonal entries, justifying the formulation used in \cref{ssec:1probleminstance}.

Lastly, note that certain permutational graph problems can be reduced to $(n-1)^2$ binary variables by fixing the starting point of each permutation. This is done by setting $x_1^1=1$, $x_1^{\alpha}=0$ for all $\alpha\neq1$, and $x_i^{i}=0$ for all $i\neq1$, which eliminates the cyclic permutation symmetry inherent in the problem, effectively reducing the search space by a factor $n$.

\section{\label{appsec:FSerrordistribution}Distribution of Finite Sampling Errors}

As quantum computing advances toward fully error-corrected systems, the only remaining source of stochastic noise will be that arising from the finite sampling of circuit measurements. In the VQA paradigm (cf.~\cref{ssec:vqas}), the number of measurement shots $n_{\mathrm{shots}}$ directly determines the precision of the estimated expectation value $\mathbb{E}[\mathbf{C}]$ relative to the true loss value $\langle\mathbf{C}\rangle$. 

For a single qubit measurement with $n_{\mathrm{shots}}$ shots, the count of “0” outcomes ($k_0$) follows a binomial distribution. According to the central limit theorem~\cite{Fischer.2011}, as ${n_{\mathrm{shots}}\rightarrow\infty}$, this distribution approaches a normal distribution with mean $\mu(k_0)=n_{\mathrm{shots}}\:p_0$ and standard deviation $\sigma(k_0)=\sqrt{n_{\mathrm{shots}}\:p_0(1-p_0)}$, where $p_0$ is the probability of measuring “0”. The corresponding standard error of the measured proportion $\hat{p}_0=\frac{k_0}{n_{\mathrm{shots}}}$ is given by:
\begin{equation}
\sigma(\hat{p}_0) = \sqrt{\frac{p_0(1-p_0)}{n_{\mathrm{shots}}}}.
\label{eq:standarderrorofmeasuredproportion}
\end{equation}
In standard VQA measurement schemes, such as with VQE and QAOA, the cost operator \eqref{eq:IsingCostFunction} is used as the measurement observable. Thus, for a parametrized quantum state $|\boldsymbol{\psi}(\boldsymbol{\theta})\rangle$, the expectation value is:
\begin{equation}
    \langle\boldsymbol{\psi}(\boldsymbol{\theta})|\mathbf{C}|\boldsymbol{\psi}(\boldsymbol{\theta})\rangle=
    \sum_{j=1}^{2^n}\:p_j\:\langle \boldsymbol{q}_j|\mathbf{C}|\boldsymbol{q}_j\rangle.
\end{equation}
Here, $p_j=\lvert\langle\boldsymbol{\psi}(\boldsymbol{\theta})|\boldsymbol{q}_j\rangle\rvert^2$ is the probability of measuring the system in the basis state $|\boldsymbol{q}_j\rangle$, which can only be estimated from measurement samples on a quantum device. The measurement counts of basis states follow a multinomial distribution, such that each estimated proportion $\hat{p}_j$ carries the standard error given in \cref{eq:standarderrorofmeasuredproportion}. 

By applying the linear approximation of error propagation and using the fact that variances of multiple independent variables add in quadrature, the statistical uncertainty of the estimated expectation value can be approximated by:
\begin{equation}
\begin{aligned}
\sigma(\mathbb{E}[\mathbf{C}])&\approx\sqrt{\sum_{j=1}^{2^n}\left(\frac{\mathrm{d}\mathbb{E}[\mathbf{C}]}{\mathrm{d}\hat{p_j}}\biggl|_{\hat{p}_j=p_j}\ \sigma(\hat{p_j})\right)^2}\\ 
&=\sqrt{\sum_{j=1}^{2^n}\left(\langle\boldsymbol{q}_j|\mathbf{C}|\boldsymbol{q}_j\rangle\right)^2 \sigma(\hat{p_j})^2}.
\end{aligned}
\label{eq:FSerrorofloss}
\end{equation}
Here, the dependence on the system size $n$ is embedded in the expectation value of the cost operator, as given by \cref{eq:IsingCostExpectation}, while the dependence on $n_{\mathrm{shots}}$ is embedded in the standard error described by \cref{eq:standarderrorofmeasuredproportion}. This behavior is empirically demonstrated in \cref{fig:FSerrors}, confirming the scaling of statistical uncertainty as a function of system size and sampling rates.

\begin{figure}
\includegraphics[scale=0.69]{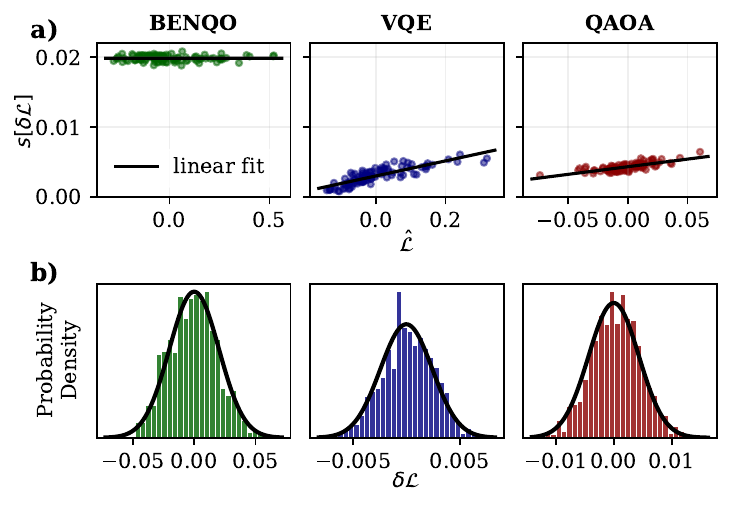}
\caption{\label{fig:FSpointerrordistribution}Distribution of finite sampling errors across loss landscape. (a) Standard deviation of the total errors from $1000$ loss evaluations with $1024$ measurement shots, at $100$ randomly selected points across the loss landscape of an $n=6$ QUBO problem instance (cf.~\cref{ssec:1probleminstance}). The relationship between the normalized loss value and the error magnitude is analyzed for three different VQAs, with linear fits indicating whether the noise behaves additively or multiplicatively. (b) Probability density histograms of $1000$ normalized errors at a single representative point, compared to the theoreticcal normal distribution with zero mean and the same standard deviation, validating the central limit theorem.}
\end{figure}

In \cref{ssec:3noisemodel}, we argue that these finite sampling errors can be reasonably modeled by adding Gaussian noise to the exact values of the loss function, assuming that this is a good-enough characterization of how statistical errors are distributed across the parameter space during loss evaluation. To illustrate this, \cref{fig:FSpointerrordistribution} presents the distribution of total normalized errors $\delta\mathcal{L}$ when sampling the loss at a randomly chosen point $\boldsymbol{\theta}\in[-2\pi,2\pi]^n$ for a system of size $n=6$ with $n_{\mathrm{shots}}=1024$ shots. These distributions are computed across the three VQAs discussed in \cref{ssec:2lossfunctions} and compared to a normal distribution with a mean of zero and the same standard deviation as the empirical data. The close visual agreement between the histogram and the theoretical Gaussian curve validates the central limit theorem, which predicts that for a sufficiently large number of shots, the binomial (or multinomial) distribution from finite sampling converges to a normal distribution.

\cref{fig:FSpointerrordistribution} also investigates the relationship between the total error magnitude $s[\delta\mathcal{L}]$ and the loss value at each evaluation point. This analysis helps determine whether finite sampling noise behaved in an additive or multiplicative manner. To quantify this, linear fits of the sample standard deviations performed, yielding values for the slope $m$ and y-intercept $t$ of the lines: If $m\ll t$, the noise is predominantly additive, whereas if $m\gg t$, the noise exhibits a multiplicative behavior. 

For the BENQO algorithm, the y-intercept $t$ is significantly larger than the slope $m$, meaning that the error is almost entirely additive. For the other two algorithms, the error displays both additive and multiplicative components, with a larger multiplicative contribution. Their increased multiplicative component arises from the strong dependence of the statistical error \eqref{eq:FSerrorofloss} on $\langle\mathbf{C}\rangle$. BENQO, however, almost entirely looses this dependency due to its implicit measurement strategy~\cite{Meli.2023}. 

Despite these differences, additive errors tend to distort the loss landscape more severely than multiplicative ones~\cite{Barligea.2024}. Therefore, when modeling total finite sampling-induced errors, a worst-case scenario can always be represented by simply adding Gaussian noise to the loss function. This justifies the noise model used in our methodology (cf.~\cref{ssec:3noisemodel}).

\section{\label{appsec:optimizerranking}Ranking of Tested Optimizers}

\begin{figure}
\includegraphics[scale=0.66]{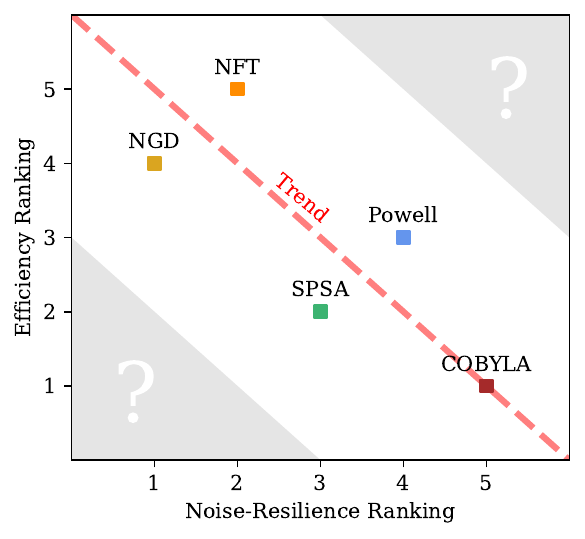}
\caption{\label{fig:optranking}Performance ranking of classical optimizers. The relationship between each optimizer's noise resilience and its efficiency ranking, based on measured mean runtimes in the same noisy experiments, is shown. The dashed diagonal line suggests a potential trend: optimizers requiring more function evaluations tend to exhibit higher noise resilience.}
\end{figure}

Inspired by Singh et al.~\cite{Singh.2023}, we present a performance ranking of the tested classical optimizers in \cref{fig:optranking}. This comparative ranking examines the trade-off between solver efficiency and noise resilience, where efficiency was measured by the mean runtimes across all experiments, directly correlating with the number of loss function calls. A clear trend emerges: optimizers with greater noise resilience tend to require more function evaluations to converge, forming an inherent trade-off between these two critical performance measures. This suggests that more noise-tolerant optimizers compensate for the lost precision by requiring additional function calls, thereby sacrificing efficiency to maintain stability in noisy settings. 

Certain optimizers -- CG, SLSQP, and BFGS -- were excluded from this ranking due to their poor performance across all tested noise levels. Their failure likely stems from their reliance on precise gradient calculations, which become very unreliable in the presence of noise. This supports previous findings~\cite{PellowJarman.2021,Singh.2023}, indicating that gradient-based optimizers, while efficient in ideal conditions, struggle in noisy settings, whereas gradient-free strategies prove significantly more robust.

Lastly, note that while this ranking provides a comparative assessment, it is important to acknowledge its limitations. The approach, inspired by Singh et al.~\cite{Singh.2023}, assigns discrete ranks rather than directly plotting continuous performance metrics. As a result, it does not fully capture quantitative performance differences between optimizers. A more precise analysis would involve directly plotting the number of function calls $n_{\text{calls}}$ against the noise resilience threshold $\sigma^*$ \eqref{eq:sigmastar}, allowing for a more detailed examination of the performance trade-offs.

\section{\label{appsec:optimalitydistribution}Optimality Distribution over Classical Solution Space}

\begin{figure}
\includegraphics[scale=0.69]{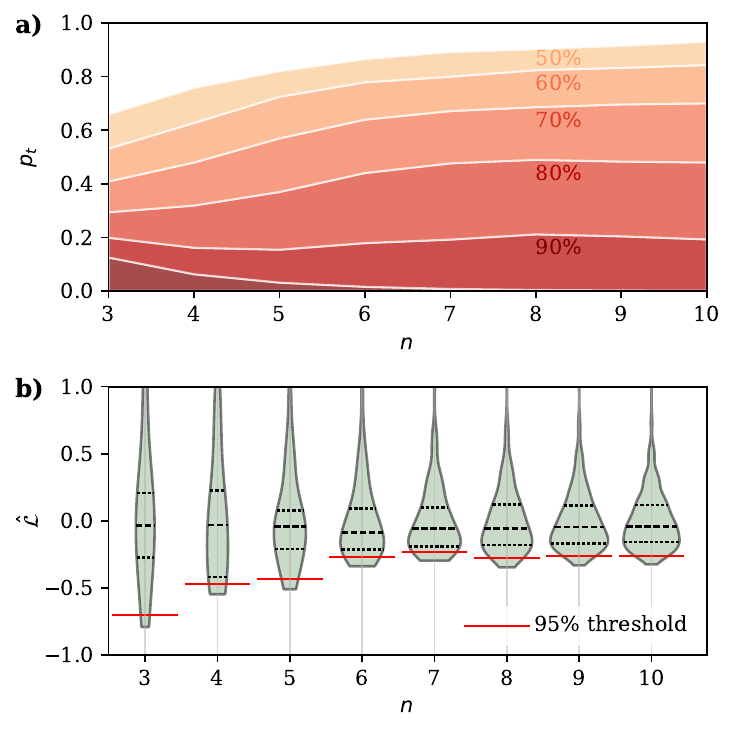}
\caption{\label{fig:solutionspace} Distribution of solution optimality. (a) The percentage of solutions $p_t$ that achieve a loss value within a relative threshold $t$ of the optimal loss value, across varying system sizes $n$, is depicted as contours. (b) Violin plots illustrate the distribution of all $2^n$ possible classical solution states within the normalized loss landscape $\hat{\mathcal{L}}$. The loss value corresponding to the $95\%$ optimality threshold is marked for each $n$, providing insight into the spread and concentration of near-optimal solutions.}
\end{figure}

This section examines the distribution of all possible solutions within the loss landscape, assessing the significance of solutions falling within the $99\%$, and $95\%$ thresholds. \cref{fig:solutionspace} depicts the results of this investigation. The data for these plots is derived from the same 100 problem instances used in our optimizer studies.

The upper panel of \cref{fig:solutionspace} presents the total percentage of solutions $p_t$ that achieve a loss value within different relative thresholds $t$  of the optimal loss (cf.~\cref{eq:optimalsolutionprobability}) across all $2^n$ possible classical solution states, i.e., bit strings. Note that this corresponds to the success probabilities of random guessing. Interestingly, this percentage appears to converge for system sizes beyond $7$ or $8$. More importantly, nearly $20\%$ of all possible solution states still yield loss value within $90\%$ of the optimum, indicating that the solution space is highly structured rather than uniformly random. 

The lower panel of \cref{fig:solutionspace} shows the corresponding normalized loss distribution of all solution state, which also seems to converge to a fixed shape, Especially, as system size grows, this distribution exhibits a left-skewed shape, with a higher concentration of solutions near the minimum energy. This deviation from a normal distribution explains why solutions above high $t$ values are more common than one might expect from a purely random search. This tendency of solutions to cluster closer to the global minimum than to the global maximum directly influences the behavior of $\hat{p}_{99\%}$, and $\hat{p}_{95\%}$ in our experiments.  

\section{\label{appsec:fitanalysis}Fit Analysis and Parameters for Observed Noise Resilience Decay}
\cref{fig:fittingexperiments} presents the fitted decay functions (cf.~\cref{eq:fit_exp,eq:fit_pl,eq:fit_log}) for the declining noise resilience threshold $\sigma^*(n)$ with respect to $\hat{p}_{\mathrm{opt}}$ across all tested optimizers. To assess the quality of these fits, we compute the mean squared error (MSE)
\begin{equation}
    d_{\mathrm{mse}} = \frac{1}{8} \sum_{n=3}^{10}\ \left [\sigma^*(n)-f(n)\right ]^2.
\label{eq:3_0MSE}
\end{equation}
The resulting MSE values, shown in \cref{tab:MSEofScalingFits}, serve as a measure of the goodness-of-fit for each optimizer and decay model. Unfortunately, these values remain inconclusive, as no single decay model consistently provides the best fit across all optimizers, with the lowest MSE dependent on the optimizer. Consequently, these results do not allow for a definitive statement on the precise nature of the decay of $\sigma^*(n)$ with increasing system size.

\begin{table}
\begin{tabular}{l|c|c|c|c|c}
\multicolumn{6}{c}{}\\[0.2ex]
$\hat{p}_{\mathrm{opt}}$            & NGD          & NFT          & COBYLA       & Powell       & SPSA         \\ \hline
exp & \scalebox{.9}[1.0]{\textbf{\num[tight-spacing=true]{1.0e-4}}} & \num[tight-spacing=true]{1.1e-3} & \num[tight-spacing=true]{5.0e-6} & \num[tight-spacing=true]{2.8e-5} & \num[tight-spacing=true]{1.6e-4} \\
pl &
\num[tight-spacing=true]{6.7e-4} &
  \num[tight-spacing=true]{1.3e-4} &
  \scalebox{.9}[1.0]{\textbf{\num[tight-spacing=true]{4.1e-7}}} &
  \scalebox{.9}[1.0]{\textbf{\num[tight-spacing=true]{5.6e-6}}} &
  \num[tight-spacing=true]{1.5e-5} \\
log &
  \num[tight-spacing=true]{1.6e-3} &
  \scalebox{.9}[1.0]{\textbf{\num[tight-spacing=true]{8.4e-5}}} &
  \num[tight-spacing=true]{8.8e-7} &
  \num[tight-spacing=true]{5.5e-5} &
  \scalebox{.9}[1.0]{\textbf{\num[tight-spacing=true]{1.3e-5}}}\\
  
\multicolumn{6}{c}{}\\[0.2ex]
$\hat{p}_{95\%}$            & NGD          & NFT          & COBYLA       & Powell       & SPSA         \\ \hline
exp & \num[tight-spacing=true]{3.5e-3} & \num[tight-spacing=true]{6.4e-3} & \num[tight-spacing=true]{1.4e-5} & \num[tight-spacing=true]{2.0e-4} & \num[tight-spacing=true]{1.1e-3} \\
pl &
\num[tight-spacing=true]{1.4e-3} &
  \num[tight-spacing=true]{2.6e-3} &
  \num[tight-spacing=true]{1.8e-6} &
  \num[tight-spacing=true]{4.0e-5} &
  \num[tight-spacing=true]{4.7e-4} \\
log &
  \scalebox{.9}[1.0]{\textbf{\num[tight-spacing=true]{8.3e-4}}} &
  \scalebox{.9}[1.0]{\textbf{\num[tight-spacing=true]{1.1e-3}}} &
  \scalebox{.9}[1.0]{\textbf{\num[tight-spacing=true]{7.4e-7}}} &
  \scalebox{.9}[1.0]{\textbf{\num[tight-spacing=true]{1.9e-5}}} &
  \scalebox{.9}[1.0]{\textbf{\num[tight-spacing=true]{2.3e-4}}}

\end{tabular}
\caption{\label{tab:MSEofScalingFits}MSE of decay function fits. The MSE values for each optimizer's noise resilience threshold $\sigma^*(n)$ w.r.t.~both $\hat{p}_{\mathrm{opt}}$ (upper) and $\hat{p}_{95\%}$ (lower) are computed for three decay models, exponential (exp), power-law (pl), and logarithmic (log), following \cref{eq:fit_exp,eq:fit_pl,eq:fit_log}. The lowest MSE for each optimizer is marked for better comparability.}
\end{table}

\begin{table*}
\begin{tabular}{ll|ll|ll|ll|ll|ll}
\multicolumn{2}{c|}{for $\hat{p}_{\mathrm{opt}}$}
   &
  \multicolumn{2}{c|}{NGD} &
  \multicolumn{2}{c|}{NFT} &
  \multicolumn{2}{c|}{COBYLA} &
  \multicolumn{2}{c|}{Powell} &
  \multicolumn{2}{c}{SPSA} \\ \cline{3-12} 
&
   &
  \multicolumn{1}{c|}{$k^*$} &
  \multicolumn{1}{c|}{$\gamma^*$} &
  \multicolumn{1}{c|}{$k^*$} &
  \multicolumn{1}{c|}{$\gamma^*$} &
  \multicolumn{1}{c|}{$k^*$} &
  \multicolumn{1}{c|}{$\gamma^*$} &
  \multicolumn{1}{c|}{$k^*$} &
  \multicolumn{1}{c|}{$\gamma^*$} &
  \multicolumn{1}{c|}{$k^*$} &
  \multicolumn{1}{c}{$\gamma^*$} \\ \hline
\multicolumn{1}{l|}{\multirow{2}{*}{exp}} &
  $\sigma^*$ &
  \multicolumn{1}{l|}{\num{3.1\pm0.2}} &
  \num{0.52\pm0.02} &
  \multicolumn{1}{l|}{\num{11\pm3}} &
  \num{0.83\pm0.07} &
  \multicolumn{1}{l|}{\num{1.7\pm0.3}} &
  \num{0.98\pm0.06} &
  \multicolumn{1}{l|}{\num{1.8\pm0.2}} &
  \num{0.66\pm0.03} &
  \multicolumn{1}{l|}{\num{2.5\pm0.5}} &
  \num{0.69\pm0.06} \\
\multicolumn{1}{l|}{} &
  $\epsilon^*$ &
  \multicolumn{1}{l|}{\num{7.3 \pm 0.7}} &
  \num{0.37\pm 0.03} &
  \multicolumn{1}{l|}{\num{18\pm3}} &
  \num{0.58\pm0.05} &
  \multicolumn{1}{l|}{\num{2.8\pm0.4}} &
  \num{0.73\pm0.05} &
  \multicolumn{1}{l|}{\num{3.5\pm0.2}} &
  \num{0.46\pm0.02} &
  \multicolumn{1}{l|}{\num{4.3\pm0.7}} &
  \num{0.46\pm0.04} \\ \hline
\multicolumn{1}{l|}{\multirow{2}{*}{pl}} &
  $\sigma^*$ &
  \multicolumn{1}{l|}{\num{8\pm2}} &
  \num{2.3\pm0.2} &
  \multicolumn{1}{l|}{\num{32\pm3}} &
  \num{3.21\pm0.08} &
  \multicolumn{1}{l|}{\num{5.1\pm0.4}} &
  \num{3.68\pm0.06} &
  \multicolumn{1}{l|}{\num{4.9\pm0.3}} &
  \num{2.71\pm0.05} &
  \multicolumn{1}{l|}{\num{6.8\pm0.5}} &
  \num{2.78\pm0.06} \\
\multicolumn{1}{l|}{} &
  $\epsilon^*$ &
  \multicolumn{1}{l|}{\num{16 \pm 4}} &
  \num{1.7\pm 0.2} &
  \multicolumn{1}{l|}{\num{49\pm2}} &
  \num{2.44\pm0.04} &
  \multicolumn{1}{l|}{\num{7.7\pm0.3}} &
  \num{2.91\pm0.03} &
  \multicolumn{1}{l|}{\num{9\pm1}} &
  \num{2.1\pm0.1} &
  \multicolumn{1}{l|}{\num{10.6\pm0.7}} &
  \num{2.03\pm0.05} \\ \hline
\multicolumn{1}{l|}{\multirow{2}{*}{log}} &
  $\sigma^*$ &
  \multicolumn{1}{l|}{\num{0.91\pm0.08}} &
  \num{3.2\pm0.3} &
  \multicolumn{1}{l|}{\num{1.43\pm0.02}} &
  \num{4.38\pm0.08} &
  \multicolumn{1}{l|}{\num{0.142\pm0.003}} &
  \num{4.9\pm0.1} &
  \multicolumn{1}{l|}{\num{0.36\pm0.02}} &
  \num{3.8\pm0.2} &
  \multicolumn{1}{l|}{\num{0.47\pm0.01}} &
  \num{3.85\pm0.08} \\
\multicolumn{1}{l|}{} &
  $\epsilon^*$ &
  \multicolumn{1}{l|}{\num{3.1 \pm 0.4}} &
  \num{2.5 \pm 0.4} &
  \multicolumn{1}{l|}{\num{4.7\pm0.1}} &
  \num{3.4\pm0.1} &
  \multicolumn{1}{l|}{\num{0.46\pm0.01}} &
  \num{4.0\pm0.1} &
  \multicolumn{1}{l|}{\num{1.20\pm0.08}} &
  \num{3.0\pm0.2} &
  \multicolumn{1}{l|}{\num{1.53\pm0.05}} &
  \num{2.9\pm0.1} \\
  
  \multicolumn{12}{c}{} \\[0.1ex] \\

\multicolumn{2}{c|}{for $\hat{p}_{95\%}$}
   &
  \multicolumn{2}{c|}{NGD} &
  \multicolumn{2}{c|}{NFT} &
  \multicolumn{2}{c|}{COBYLA} &
  \multicolumn{2}{c|}{Powell} &
  \multicolumn{2}{c}{SPSA} \\ \cline{3-12} 
&
   &
  \multicolumn{1}{c|}{$k^*$} &
  \multicolumn{1}{c|}{$\gamma^*$} &
  \multicolumn{1}{c|}{$k^*$} &
  \multicolumn{1}{c|}{$\gamma^*$} &
  \multicolumn{1}{c|}{$k^*$} &
  \multicolumn{1}{c|}{$\gamma^*$} &
  \multicolumn{1}{c|}{$k^*$} &
  \multicolumn{1}{c|}{$\gamma^*$} &
  \multicolumn{1}{c|}{$k^*$} &
  \multicolumn{1}{c}{$\gamma^*$} \\ \hline
\multicolumn{1}{l|}{\multirow{2}{*}{exp}} &
  $\sigma^*$ &
  \multicolumn{1}{l|}{\num{1.4\pm0.3}} &
  \num{0.30\pm0.06} &
  \multicolumn{1}{l|}{\num{5\pm2}} &
  \num{0.6\pm0.1} &
  \multicolumn{1}{l|}{\num{0.8\pm0.2}} &
  \num{0.73\pm0.07} &
  \multicolumn{1}{l|}{\num{1.1\pm0.2}} &
  \num{0.52\pm0.06} &
  \multicolumn{1}{l|}{\num{1.2\pm0.4}} &
  \num{0.5\pm0.1} \\
\multicolumn{1}{l|}{} &
  $\epsilon^*$ &
  \multicolumn{1}{l|}{\num{2.9 \pm 0.6}} &
  \num{0.13\pm 0.04} &
  \multicolumn{1}{l|}{\num{7\pm2}} &
  \num{0.28\pm0.06} &
  \multicolumn{1}{l|}{\num{1.3\pm0.2}} &
  \num{0.48\pm0.04} &
  \multicolumn{1}{l|}{\num{2.0\pm0.3}} &
  \num{0.30\pm0.04} &
  \multicolumn{1}{l|}{\num{1.7\pm0.5}} &
  \num{0.19\pm0.06} \\ \hline
\multicolumn{1}{l|}{\multirow{2}{*}{pl}} &
  $\sigma^*$ &
  \multicolumn{1}{l|}{\num{3.1\pm0.7}} &
  \num{1.5\pm0.2} &
  \multicolumn{1}{l|}{\num{12\pm3}} &
  \num{2.3\pm0.2} &
  \multicolumn{1}{l|}{\num{2.1\pm0.2}} &
  \num{2.89\pm0.09} &
  \multicolumn{1}{l|}{\num{2.8\pm0.4}} &
  \num{2.2\pm0.1} &
  \multicolumn{1}{l|}{\num{2.8\pm0.9}} &
  \num{2.0\pm0.2} \\
\multicolumn{1}{l|}{} &
  $\epsilon^*$ &
  \multicolumn{1}{l|}{\num{5 \pm 1}} &
  \num{0.7\pm 0.1} &
  \multicolumn{1}{l|}{\num{15\pm4}} &
  \num{1.4\pm0.2} &
  \multicolumn{1}{l|}{\num{3.2\pm0.2}} &
  \num{2.12\pm0.05} &
  \multicolumn{1}{l|}{\num{4.4\pm0.5}} &
  \num{1.48\pm0.09} &
  \multicolumn{1}{l|}{\num{3\pm1}} &
  \num{1.1\pm0.2} \\ \hline
\multicolumn{1}{l|}{\multirow{2}{*}{log}} &
  $\sigma^*$ &
  \multicolumn{1}{l|}{\num{0.77\pm0.05}} &
  \num{2.2\pm0.2} &
  \multicolumn{1}{l|}{\num{1.27\pm0.07}} &
  \num{3.3\pm0.2} &
  \multicolumn{1}{l|}{\num{0.130\pm0.002}} &
  \num{3.99\pm0.07} &
  \multicolumn{1}{l|}{\num{0.34\pm0.01}} &
  \num{3.2\pm0.1} &
  \multicolumn{1}{l|}{\num{0.41\pm0.03}} &
  \num{2.9\pm0.2} \\
\multicolumn{1}{l|}{} &
  $\epsilon^*$ &
  \multicolumn{1}{l|}{\num{2.5 \pm 0.2}} &
  \num{1.2 \pm 0.2} &
  \multicolumn{1}{l|}{\num{4.0\pm0.3}} &
  \num{2.2\pm0.2} &
  \multicolumn{1}{l|}{\num{0.43\pm0.01}} &
  \num{3.06\pm0.09} &
  \multicolumn{1}{l|}{\num{1.09\pm0.04}} &
  \num{2.2\pm0.1} &
  \multicolumn{1}{l|}{\num{1.3\pm0.1}} &
  \num{1.7\pm0.3}
\end{tabular}
\caption{\label{tab:fitparameters}Final parameters of decay function fits. The optimized parameters $k^*$ and $\gamma^*$ for the exponential (exp), the power-law (pl) and logarithmic (log) decay functions (cf.~\cref{eq:fit_exp,eq:fit_pl,eq:fit_log}) are listed for the measured noise resilience thresholds $\sigma^*(n)$~\eqref{eq:sigmastar} and $\epsilon^*(n)$~\eqref{eq:sigma2rae}. Results are provided for all tested optimizers and both solvability measures $\hat{p}_{\mathrm{opt}}$ (upper part) based on \cref{fig:fittingexperiments} and $\hat{p}_{95\%}$ (lower part).}
\end{table*}
Finally, for reproducibility, \cref{tab:fitparameters} provides the final parameters of all fitting experiments shown in \cref{fig:fittingexperiments}. 

\newpage
\bibliography{library}

\end{document}